\documentclass[12pt]{article}
\pdfoutput=1
\usepackage{amsmath}
\usepackage{amsthm}
\usepackage{amscd}
\usepackage{amsfonts}
\usepackage[psamsfonts]{amssymb}
\usepackage{graphicx}
\usepackage{hyperref}

\numberwithin{equation}{section}

\makeatletter
\renewcommand\section{\@startsection {section}{1}{\z@}%
                  {-3.5ex \@plus -1ex \@minus -.2ex}
                  {2.3ex \@plus.2ex}%
                  {\normalfont\large\bfseries}}
\renewcommand\subsection{\@startsection{subsection}{2}{\z@}%
                  {-3.25ex\@plus -1ex \@minus -.2ex}%
                  {1.5ex \@plus .2ex}%
                  {\normalfont\bfseries}}
\renewcommand\subsubsection{\@startsection{subsubsection}{3}{\z@}%
                  {-3.25ex\@plus -1ex \@minus -.2ex}%
                  {1.5ex \@plus .2ex}%
                  {\normalfont\itshape}}
\makeatother

\begin{document}
\begin{titlepage}
\begin{flushright}
TAUP-2943/12
\end{flushright}
\vspace{0.2in}
\begin{center}{\LARGE{\bf Schr\"odinger Holography \\
with and without Hyperscaling Violation } }\\ 
\vspace{0.5in}
 
{\large \bf Bom Soo Kim}\\ ~\\ \vspace{-0.1in}
\mbox{\normalsize {Raymond and Beverly Sackler School of Physics and Astronomy,}}\\ 
\mbox{\normalsize   Tel Aviv University, 69978 Tel Aviv, Israel} \\ 
\mbox{\normalsize  \it  bskim@post.tau.ac.il}
\end{center}
\bigskip 
\vspace{0.5in}

\begin{abstract}
We study the properties of the Schr\"odinger-type non-relativistic holography for general dynamical 
exponent $z$ with and without hyperscaling violation exponent $\theta$. 
The scalar correlation function has a more general form due to general $z$ as well as the presence of $\theta$, whose 
effects also modify the scaling dimension of the scalar operator. 
We propose a prescription for minimal surfaces of this ``codimension 2 holography," 
and demonstrate the $(d-1)$ dimensional area law for the entanglement entropy from $(d+3)$ dimensional 
Schr\"odinger backgrounds. Surprisingly, the area law is violated for $d+1 < z< d+2 $, even  
{\it without} hyperscaling violation, which interpolates between the logarithmic violation and  
extensive volume dependence of entanglement entropy. 
Similar violations are also found in the presence of the hyperscaling violation. 
Their dual field theories are expected to have novel phases for the parameter range, including Fermi surface. 
We also analyze string theory embeddings using non-relativistic branes.
\end{abstract}
\end{titlepage}

\tableofcontents

\section{Introduction and Summary}     \label{sec:introduction} 

AdS /CFT (Anti-de Sitter space / Conformal Field Theory) correspondence provides a unique 
analytic approach to strongly coupled field theory \cite{Maldacena:1997re}\cite{Aharony:1999ti}. 
This tool has been extended to its application to the non-relativistic setup with dynamical exponent $z \neq 1$ 
in \cite{son}\cite{goldberger} in the context of Schr\"odinger holography and 
in \cite{Kachru:2008yh} for Lifshitz spacetime. 
Along the line of holographic application to condensed matter physics, the major players have been AdS$_4$ with 
dynamical exponent $z=1$ as well as Lifshitz spacetime and its generalizations. 
While it is very interesting to consider Schr\"odinger-type theories, the progress has been slow due 
to their technical difficulties. We would like to close some of the gaps existing in the literature 
following recent progress \cite{Ogawa:2011bz}\cite{Huijse:2011ef}\cite{Dong:2012se}.  

For the holographic applications to the condensed matter system, it becomes clear that we need to consider 
more general classes of metrics than those with the asymptotic AdS boundaries. 
More drastically, one can consider only small part of the `AdS' space by {\it cutting out} the ultraviolet 
part. There has been several different approaches \cite{CGKKM}\cite{rgflow}. 
Let us consider the Effective Holographic approach (EHT) initiated in \cite{CGKKM} following the Effective 
Field theory approach. In \cite{CGKKM}, a general set of metric is constructed 
in the context of Einstein-Maxwell-Dilaton (EMD) theories with a general dilaton coupling to the Maxwell field 
and a dilaton potential. These coupling and potential have exponential forms with two parameters $\beta$ and $\gamma$. 
The parameter spaces are constrained by various physical conditions in IR, 
such as regularity condition at the horizon, Gubser's bound and well defined fluctuation problems.
The resulting solutions are extensively analyzed for the thermodynamic and transport properties in \cite{CGKKM}%
\footnote{See \cite{GKMReview} for the simplified discussions of \cite{CGKKM}.} 
and for the fermionic spectral functions in \cite{IKNT}. 

The full set of near extremal solution for $d=2$%
\footnote{General $d$ dimensional metric is also presented in \cite{CGKKM}. 
And its thermodynamic and transport properties are also analyzed there.}
 is obtained in \cite{CGKKM}
  	\begin{align}    \label{EMDsolution}
  		&S =\int d^{p+1}x~\sqrt{-g}\left[R- \frac{e^{\gamma\phi}}{4}F_{\mu\nu}F^{\mu\nu}
  		-\frac{1}{2}(\partial\phi)^2-2\Lambda e^{-\delta\phi}\right]  \;, \nonumber  \\ 
		&d s^2 = -r(r-2m) r^{-4\frac{\gamma(\gamma-\delta)}{wu}} dt^2 
		+\frac{ e^{\delta\phi}  dr^2}{ -w\Lambda r(r-2m)} 
		+ r^{2\frac{(\gamma-\delta)^2}{wu}}  \left( dx^2 + dy^2 \right) \;,   \\
		&e^{\phi} = e^{\phi_0}r^{-4(\gamma-\delta)/(wu)}\;, \quad 
		\mathcal A_t= \Phi + 2\sqrt{-v/(wu)}e^{-\frac{\gamma}{2} \phi_0}r  \;,  \nonumber \\
		&wu=3\gamma^2-\delta^2 - 2\gamma \delta + 4 \;, 
		\quad u = \gamma^2- \gamma \delta +2 \;, \quad v=\delta^2 - \gamma \delta -2 \;.  \nonumber
	\end{align}
This metric was constructed to get a general scaling solution in IR, and physical properties 
such as entropy, energy and conductivities show power law behaviors, characteristic features of 
scaling invariant theories. An important difference is that the hyperscaling is violated.   
Lifshitz solution can be obtained with $\gamma = -\sqrt{4/(z-1)}$ and $\delta=0$. 
This solution provides the most general IR asymptotics at finite density with a single gauge field $\mathcal A$ 
and a dilaton field $\phi$, and is embedded in higher dimensional AdS or Lifshitz spacetime 
\cite{Gouteraux:2011ce}.%
\footnote{Hyperscaling violation is first mentioned in the holographic context in \cite{Gouteraux:2011ce}. 
We thank  to Elias Kiritsis and Subir Sachdev for the discussions on the hyperscaling violation 
during the conference ``Black Hole Answers for Condensed Matter Questions'' held in Leiden, June 14 - 17, 2011.
} 
This solution can be understood along the line of developments for EMD theories \cite{EMDTheories}\cite{Harrison:2012vy}.

Hyperscaling is a property of the free energy based its naive dimension \cite{SachdevBook}. 
For the theories with hyperscaling, the entropy behaves as $S \sim T^{d/z}$, where $T, d$ and $z$ are temperature, 
number of spatial dimensions and dynamical exponent. Hyperscaling was shown to be violated by random-field 
fluctuations, which dominate over thermal fluctuations at long length scale, and cause the scale of 
free energy to grow with modified scaling \cite{Fisher}. 
In holographic context, the hyperscaling violation exponent $\theta$ is related to the transformation of 
the proper distance (see \ref{metricTransformHS}), and thus its non-invariance implies the violation of 
hyperscaling of the dual field theory \cite{Huijse:2011ef}. 
Then, we have modified relation between the entropy and temperature as $S \sim T^{(d-\theta)/z}$.

Recently, Lifshitz-type theories with hyperscaling violation is proposed in more general setup 
\cite{Dong:2012se}\cite{Huijse:2011ef}\cite{Ogawa:2011bz} without referring to particular matter contents. 
To have physically reasonable theories, null energy condition is imposed. The metric is proposed as 
	\begin{align} \label{LifshitzMetric}
		ds^2 = r^{-2 + 2  \theta/d} \left( - \beta r^{-2(z-1)} dt^2 + \sum_{i=1}^{d} d x_i^2 + dr^2  \right) \;, 
	\end{align}
where $z$ and $\theta$ are the dynamical and hyperscaling violation exponents, respectively.   
The EMD solution given in (\ref{EMDsolution}) can be rewritten in this form using the coordinate transform 
$ r \rightarrow r^a$ with $ a = - \frac{wu}{\gamma^2 -\delta^2}$. 
For vanishing non-extremality parameter $m=0$, the dynamical and hyperscaling exponents are given in terms 
of $\gamma, \delta$ for $d=2$ as  	
	\begin{align}
		\theta = \frac{4 \delta }{\gamma + \delta} \;, \quad 
		z = 1 + \frac{2\delta}{\gamma + \delta } + \frac{4}{\gamma^2 -\delta^2} \;. 
	\end{align}
Thus we can translate all the results of \cite{CGKKM} in terms of $z$ and $\theta$. Again, we check that 
$\delta=0$ is a Lifshitz solution. 

In \cite{Dong:2012se}, the basic properties of the Lifshitz systems with metric (\ref{LifshitzMetric}) are 
analyzed. The correlation function of scalar operators signals that the scaling dimension of the dual operator 
is shifted by the exponent $\theta$. Furthermore, the authors of \cite{Dong:2012se} analyzed the 
holographic entanglement entropy to find the existence of novel phases for some range of the parameter 
$\theta$, where entanglement entropy violates the area law. 
These phases are identified to interpolate between that of a Fermi surface and that 
exhibit extensive entanglement entropy. 

This brings us to another important subject in holographic application to condensed matter: 
identifying and understanding Fermi surfaces. 
Fermi surface in holographic context was first realized and studies in Reissner-Nordstr\"om black hole 
\cite{RNFermiSurface}. These are known to violate the Luttinger theorem, which relates the area of the 
Fermi surface to the total charge. (See {\it e.g.} \cite{LuttingerFS} for satisfying the theorem, and also 
\cite{Allais:2012ye}, along the line of ``electron star'' \cite{ElectronStar}.) 
It was argued in \cite{Sachdev:2010um} that holographic theories can only access gauge-invariant 
parts of Fermi surfaces and thus the rests are hidden Fermi surfaces of 'fractionalized' fermions 
carrying gauge charges. 

A very nice resolution is realized in the context of the holographic entanglement entropy \cite{HoloEntanglement}. 
Entanglement entropy has become a new useful tool to understand different phases of field theory, 
differentiating the fermionic models compared to bosonic ones \cite{Eisert:2008ur}. 
The authors of \cite{Ogawa:2011bz}  proposed a new definition of the systems with Fermi surface as 
``the entanglement entropies of the system with Fermi surface show the logarithmic violation of the area law.'' 
This property has been shown for non-Fermi liquids as well as free fermion systems \cite{Entanglement}. 
Using the holographic entanglement entropy, 
new novel phases were found in \cite{Dong:2012se} for the Lifshitz theories with hyperscaling violation.%
\footnote{See further developments for holographic entanglement entropy 
in this context \cite{ProgEntangle}.}  

Inspired by these recent developments, we investigate Schr\"odinger-type systems with Galilean 
invariance for arbitrary dynamical exponent $z$ with and without hyperscaling violation exponent $\theta$. 
Our analysis is parallel to that of \cite{Dong:2012se} and find several surprises as well as similar properties.   

The metric for $3+d$ dimensional Schr\"odinger-type theories is given by   
	\begin{align} \label{zeroTMetric}
		ds^2 = r^{-2 + 2  \theta/D} \left( - \beta r^{-2(z-1)} dt^2 - 2 dt d\xi + \sum_{i=1}^{d} d x_i^2 + dr^2  \right) \;, 
	\end{align}
where $\theta$ signifies the hyperscaling violation with a factor $D$, which we leave undetermined, 
and $z$ is a dynamical exponent. 
Note that there exists an off-diagonal component of the metric, which is crucial to maintain the Galilean boost. 
For $z=2$ and $\theta=0$, the symmetry is extended to non-relativistic conformal symmetry including special conformal 
invariance. Before going into detailed analysis, let us summarize main results here. 

The basic properties of Schr\"odinger-like systems with metric (\ref{zeroTMetric}) are analyzed in 
\S \ref{sec:basics}. We constrain the parameters $(\theta, z, d, D)$ of the theory using 
null energy condition. The result is given in (\ref{nullECondition}) and the allowed regions are depicted 
in figure \ref{fig:AllowedRegions}. By adapting effective holographic approach, we calculate the 
stress-energy tensor on the hypersurface located at finite $r=r_c$, whose conformal weight is shifted by 
$-(d+1)\theta/D$. This result is in (\ref{HSET}). Similar shifting is also observed in correlation functions.   

In \S \ref{sec:WKBPropagator}, we evaluate the geodesic distances of a particle with mass $m$ using semiclassical 
approximation. This can be identified as a massive propagator. In the semiclassical limit, the propagator has 
exponentially decaying behavior with non-trivial $\theta$ dependence (\ref{staticPropagator}) 
(\ref{stationaryLxiPropagator}) (\ref{timelikePropagatorSmallP2}) (\ref{timelikePit=0case}) 
(\ref{timelikePropagatorSmallP}). As $\theta \rightarrow 0$, the propagator shows power-law behavior (\ref{scalingPropagator}).  

In section \S \ref{sec:correlationFunction}, we couple a scalar field in the background and analyze 
full scalar correlation functions beyond the semiclassical approximation. 
From the analysis of many different cases, we find that the scaling dimension of the dual scalar  
operator is modified as 
	\begin{align}
		\Delta_+ = \Delta_{+, \theta=0} -\frac{(d+1)\theta}{D} \;,   	
	\end{align}
and $\Delta_- =0 $, when $m^2$, mass of the scalar field, and $M^2$, the eigenvalue along the $\xi$ coordinate, 
do not contribute to the scaling dimensions. 
For the case with the contribution of $M^2$, see (\ref{generalScalingDimension}), where there are 
two independent nonzero dimensions $\Delta_+$ and $\Delta_-$.   
In general, the momentum space correlation function has the form  
	\begin{align}      
		G(\omega, \vec k) \sim \left(\vec k^2 - 2 M \omega + M^2 + m^2 \right)^{(\Delta_+ - \Delta_-)/2} \;,
	\end{align} 
and the general scalar correlation function in position space is 
	\begin{align}      
		\langle\mathcal O(t', \vec x')\mathcal O(t, \vec x)\rangle \sim
		\frac{\theta(t'- t)}{|t'-t|^{\Delta_+ + (d+1)\theta / 2D}} 
		e^{i M \frac{|\vec x' - \vec x|^2}{2 |t'- t|} - i \frac{M^2 + m^2}{2 M} |t'- t|}  \;.
	\end{align} 
All the examples worked out in \S \ref{sec:corrHSV} and \S \ref{sec:corrHSVAdSLC} satisfy this relation. 
We observe two new features. First, both the power dependence $|t'-t|^{\Delta_+ + (d+1)\theta / 2D}$ and $\Delta_+$
depend on the hyperscaling violation exponent $\theta$. For $\theta=0$, this reduces to 
$|t'-t|^{\Delta_{+, \theta =0}}$. 
Second, the exponent is modified by 
$\frac{M^2 + m^2}{2 M} |t'- t|$. This modification comes from the effects of the general dynamical 
exponent $z$ as well as $\theta$. 

Schr\"odinger holography has a unique feature : the holographic correspondence is ``codimension 2" meaning 
that $(d+3)$ dimensional Schr\"odinger background in gravity side corresponds to $(d+1)$ dimensional field theory.  
There exist a spectator direction $\xi$, in addition to the radial coordinate $r$ representing the 
energy scale of the dual field theory. 
We provide a prescription for the minimal surface in this ``codimension 2" holography by treating the $\xi$ 
coordinate special. This gives the same result both for static and stationary cases, 
which are explained in \S \ref{sec:EntanglementEntropy}. 
The direction $\xi$ turns out to be crucial for our prescription of minimal surface. 

While the results are shown to be valid for general shape in \S \ref{sec:generalEntangleRegion}, 
we compute the entanglement entropy for the strip geometry located at $r=\epsilon$ 
	\begin{align}
		\quad -l \le x_1 \le l\;,\quad 0 \le x_i \le L\;,\;\quad i = 2,\cdots \;, d \;.
	\end{align}
We integrate over the entire possible area for the special $\xi$ coordinate and identify it with 
total `length scale' $L_\xi$ or alternatively corresponding `mass scale' $M_\xi$, which is a defining 
property of the dual field theory. Specifically, for $z=2, \theta=0$, we get 
	\begin{align}    
		\mathcal  S_{z=2, \theta=0}  
		&= \frac{(R M_{Pl})^{(d+1)}}{4  (d-1) M_\xi }  ~  \left( \left(\frac{L}{\epsilon} \right)^{d-1}
		  -  c_{z=2} ~  \left(\frac{L}{l} \right)^{d-1}  \right) \;,
	\end{align}
where $R, M_{Pl}$ are curvature scale and $(d+3)$ dimensional Plank mass. 
$M_\xi$ is the dimensionless `mass scale' associated with $\xi$ coordinate.   
$c_{z=2}$ is constant and this result is given in (\ref{entanglementEntropySchrz=2}). 
This describes the $(d-1)$ dimensional area in the entanglement entropy, computed from the $(d+3)$ 
dimensional Schr\"odinger background (\ref{zeroTMetric}). 

Surprisingly, we find the logarithmic violation of the area law even {\it without} hyperscaling violation 
$\theta=0$ for $z=d+1$ as 
	\begin{align}    
		\mathcal  S_{z=d+1, \theta=0}  &= \frac{(R M_{Pl})^{(d+1)}}{4 \beta^{1/2}} 
		\left( \frac{L^{d-1} }{M_\xi} \right) \log \left(\frac{2l}{\epsilon} \right)  \;,
	\end{align}   
which signals the presence of Fermi surface in the dual field theory according to \cite{Ogawa:2011bz}. 
This is worked out in (\ref{logEntropyz}). Note the dimensionless combination, $L^{d-1}/M_\xi$, for $z=d+1$. 
Furthermore, we also find the violation of area law in the entanglement entropy for the range 
\begin{align}
d+1 < z < d+2 \;,
\end{align} 
interpolating from the logarithmic violation for $z=d+1$ to the extensive violation for $z=d+2$ in the same section 
\S \ref{sec:novelPhasez}. Thus the dual field theory is expected to reveal novel phases for the 
parameter range including Fermi surface. 

For $\theta \neq 0$, with hyperscaling violation, we work out the general entanglement entropy as
	\begin{align}      
		\mathcal  S 
		&= \frac{(R M_{Pl})^{(d+1)} }{4 (\alpha-1)} 
		\left(  \left(\frac{\epsilon}{R_\theta}\right)^{(d+1)\theta/D} \frac{L^{d-1} L_\xi }{\epsilon^{d-z+1}} 
		-c_{\theta} ~\left(\frac{l}{R_\theta}\right)^{(d+1)\theta/D} \frac{L^{d-1} L_\xi }{l^{d-z+1}}  \right) \;,
	\end{align}   
in (\ref{entanglementEntropyHyper}). $R_\theta$ is a scale where the effects of the hyperscaling violation 
becomes strong. We also find the area law violation of entanglement entropy, interpolating from the 
logarithmic violation and the extensive violation, for the following range 
	\begin{align}
		\frac{d+1-z}{d+1}  < \frac{\theta}{D} < \frac{d+2-z}{d+1}  \;, 
	\end{align}
which is analyzed in \S \ref{sec:novelPhasetheta}. 
The allowed regions of these novel phases are depicted in figure \ref{fig:NovelPhaseAllowedRegions} 
along with the allowed regions by the null energy conditions.

While we constrain our theories with null energy condition in \S \ref{sec:basics}, it is important 
to investigate string theory construction of the theories with hyperscaling violation. 
This is done in \S \ref{sec:stringConst} based on non-relativistic D$p$ brane solutions ($p=d+1$)
\cite{topdownSchrMetric}\cite{topdownSchrMetric2} using null Melvin twist \cite{nullMelvinTwist}. 
It turns out that the hyperscaling violation exponent $\theta$ vanishes for $d=2$, 
and is negative for the range $d<4$. 

In \S \ref{sec:AdSLC}, we perform a parallel study for the AdS in light-cone (ALCF) with 
the metric (\ref{zeroTAdSinLCMetric}), which is the case with $\beta=0$ in (\ref{zeroTMetric}). 
The metric (\ref{zeroTAdSinLCMetric}) describes Schr\"odinger holography for general 
$z$ and $\theta$ with conformal Schr\"odinger symmetry for any value of $z$. 
We analyzed null energy condition and find that it is independent of $z$.  
This is different from $\beta \neq 0$, but identical to AdS case. 
The correlation functions and semiclassical propagator are analyzed in \S \ref{sec:correlationFunctionAdSLC} 
and \S \ref{semiclassicalPropagatorAdSLC}. They are simpler, but reveals similar properties compared to 
those of the $\beta\neq 0$ case. Even though the metric does not depend on $z$ explicitly, 
we find the same results on entanglement entropy in \S \ref{sec:EntanglementEntropyAdSLC} 
by extending the minimal surface prescription motivated by the $\beta \neq 0$ case. 

We conclude with future directions in \S \ref{sec:outlook}, which includes some speculations for 
the finite temperature generalizations. We provide metric properties and useful formula for the computations 
in appendix, \S \ref{sec:formula} and \S \ref{sec:EinsteinTensorApp}.

\section{Schr\"odinger background with hyperscaling violation} \label{sec:basics}

We would like to consider the $3+d$ dimensional metric given in (\ref{zeroTMetric}).  
For $\theta=0$, the metric is first considered in \cite{son}. The corresponding finite temperature generalizations 
were done in \cite{Herzog:2008wg}\cite{Maldacena:2008wh}. 
The case with $\theta=0$ and $\beta=0$ is also considered in \cite{goldberger}, whose 
finite temperature generalizations were considered in \cite{Maldacena:2008wh}\cite{Kim:2010tf}\cite{KKP}. 
We would like to concentrate on $\beta \neq 0$ here and the case with $\beta=0$ is considered 
briefly in \S \ref{sec:AdSLC}.   

The metric (\ref{zeroTMetric}) is invariant under the translations, rotations and Galilean boost, 
which has the following form 
	\begin{align}
		\vec {x}^{ \prime} =\vec{x} - \vec {v}t \;, \quad 
		\xi^{ \prime} = \xi + \frac{1}{2}(2 \vec{v} \cdot \vec{x} - {v^2}t) \;,
	\end{align}
where the vectors are $d$-dimensional vectors. 
Note that, under the scale transformation
	\begin{align}   \label{ST}
		t \rightarrow \lambda^z t \;, \quad \xi \rightarrow \lambda^{2-z} \xi \;, \quad  
		\vec x \rightarrow \lambda \vec x \;, \quad r \rightarrow \lambda r \;, 
	\end{align}
the metric (\ref{zeroTMetric}) is not invariant, but transforms covariantly 
	\begin{align}      \label{metricTransformHS}
		ds \rightarrow \lambda^{\theta/D} ds\;,
	\end{align}
which is a defining property of hyperscaling violation in holographic language. 
It is a curious fact that, under the special conformal transformation 
	\begin{align}    \label{SCT}
		\vec x' = \frac{\vec x}{1+ct} \;, \quad
		t'  = \frac{t}{1+ct} \;, \quad
		r' = \frac{r}{1+ct} \;, \quad 
		\xi' = \xi + \frac{c}{2} \frac{\vec{x}\cdot\vec{x} + r^2}{1 +  ct} \;,
	\end{align} 
the metric transforms as 
	\begin{align} \label{metricCT}
		ds \rightarrow \left( \frac{r}{1+ct}\right)^{\theta/D} ds \;, \qquad \quad  
		\beta=0 \qquad  \text{or} \qquad  z=2~ \& ~\beta \neq 0 \;.
	\end{align}
For $\theta =0$, the metric (\ref{zeroTMetric}) is conformal for $\beta=0$ with general $z$ and $\beta \neq 0 ~\&~ z=2$. 
For these cases, the metric transforms also covariantly as (\ref{metricCT}). 
The metric (\ref{zeroTMetric}) is conformally equivalent to the Schr\"odinger metric for $\beta \neq 0$. 

Using the effective holographic approach \cite{CGKKM}\cite{Dong:2012se}, we consider 
the corresponding dual field theory living at a finite radius, $r=r_c$. Thus the metric (\ref{zeroTMetric}) provides 
a good description of the dual field theory only for a certain range of $r$, presumably for $r \geq r_c$ 
anticipating the applications at the low energy regions. 
While it is plausible for the theory to flow to some fixed point at $r \gg r_c$ as pointed out in \cite{Dong:2012se} 
(further developed in \cite{Harrison:2012vy}), 
the effective holographic approach in \cite{CGKKM} constrains the low energy dynamics with various physical conditions 
available in IR, such as regularity conditions, Gubser's bound \cite{Gubser:2000nd} and well defined fluctuation problems. 
To make our story simple, we assume this is the case. Thus the warp factor $e^{2A(r)} \rightarrow R^2/r^2$ as 
$r \rightarrow 0$, where $R$ is the curvature scale. Below the crossover scale $R_\theta$, the metric has the overall 
factor $R^2/R_\theta^{2\theta/D}$. These scales are ignored during the calculations below and will be restored 
at the end of calculation using dimensional analysis.    

For this purpose, we consider more general background of the 
following form, which is invariant under the translations, rotations and Galilean boost
	\begin{align} \label{generalzeroTMetric}
		ds^2 = e^{2A(r)} \left( - \beta e^{2B(r)}  dt^2 - 2 dt d\xi + \sum_{i=1}^{d} d x_i^2 + dr^2  \right) \;, 
	\end{align} 
for the description outside the region valid for (\ref{zeroTMetric}). 

It is worthwhile to mention that $\xi$ is a special isometry direction and is known to 
provide a particle number $M$ to the dual field theory for $z=2, \theta=0$.%
\footnote{See some developments along this line in \cite{Hyun:2011qj}, where $M$ is generalized to be complex 
in the context of time dependent setup. There two time correlation function was constructed to show slow dynamics, 
power law decaying behavior, for the pure imaginary $M$.   
See also \cite{Guica:2010sw}\cite{Balasubramanian:2010uw} for different considerations on this matter.} 
If we consider a static case, the metrics (\ref{zeroTMetric}) and (\ref{generalzeroTMetric}) seem to lose contact with 
the coordinate $\xi$. 
To see the effect on physical properties of $\xi$ clearly, we also consider a stationary case.%
\footnote{The minimal surface prescription, however, gives a unique result \S \ref{sec:EntanglementEntropy}.
}
Thus, we consider the ADM form of the metric  
	\begin{align} \label{ADMzeroTMetric}
		ds^2 = e^{2A(r)} \left( - \beta e^{2B(r)}  \left( dt  + \beta^{-1} e^{-2B(r)} d\xi \right)^2 
		+  \beta^{-1} e^{-2B(r)} d\xi^2 + \sum_{i=1}^{d} d x_i^2 + dr^2  \right) \;. 
	\end{align}
We impose the condition $dt  + \beta^{-1} e^{-2B(r)} d\xi = 0$ for a stationary case when we consider 
massive propagator in \S \ref{sec:WKBPropagator} and entanglement entropy in \S \ref{sec:EntanglementEntropy}.

\subsection{Metric properties and the null energy condition}

Rather than generating a particular solution with the hyperscaling violation, we find 
some constraints on the parameters of the metric (\ref{zeroTMetric}) using null energy condition 
\cite{Ogawa:2011bz}\cite{Huijse:2011ef}\cite{Dong:2012se}.%
\footnote{In \cite{Hoyos:2010at}, null energy condition is used to rule out holographic Lifshitz backgrounds 
with $z<1$.  
}      

The Ricci tensors and scalar curvature for the metric (\ref{zeroTMetric}) are given by 
(see \S \ref{sec:EinsteinTensorApp})
	\begin{align}    \label{RicciTensor}
		R_{tt}&=\beta \frac{D^2(2+(d-2)z+2z^2 )- D((d+2) +(d+1)z) \theta + (d+1) \theta^2 }{D^2 r^{2z}} \;, \nonumber\\
		R_{ii}&= - R_{t\xi} = - \frac{(D-\theta) (D(d+2) - (d+1)\theta)}{D^2 r^2} \;,  \nonumber\\
		R_{rr}&=\frac{(d+2)( \theta-D) }{D r^2} \;, \nonumber \\ 
		\mathcal R ~~&= r^{-2\theta/D} \frac{(d+2)(\theta-D) (D(d+3) - (d+1)\theta)}{D^2} \;.
	\end{align}
The scalar curvature is $\mathcal R \propto r^{-2 \theta /D}$, which becomes constant for $\theta=0$ as expected 
from the observation that the metric (\ref{zeroTMetric}) is conformally equivalent to Schr\"odinger metric. 
Energy momentum tensor can be computed as $T_{\mu\nu} = R_{\mu\nu} - 1/2 g_{\mu\nu} \mathcal R$.%
\footnote{Explicit solution which supports this Einstein tensor is recently constructed in \cite{Perlmutter:2012he}. 
The corresponding matter contents are a massive vector field with an additional scalar field with appropriate gauge 
coupling and scalar potential described by a similar action considered in \cite{CGKKM}. 
}  

To consider various physically sensible dual field theories, 
we would like to constrain the parameters using the null energy condition 
	\begin{align}   
		T_{\mu\nu} N^\mu N^\nu \geq 0 \;,
	\end{align}   
where the null vectors satisfy $N^\mu N_\mu = 0$. 
The two independent null vectors are 
	\begin{align}   
		N^t= \frac{1}{\beta^{1/2} r^{\theta /D - z }} \;, \quad N^r= \frac{\cos (\phi )}{r^{-1 +\theta /D}} \;, 
		\quad N^i= \frac{ \sin (\phi) }{r^{-1 +\theta /D}}  \;, 
	\end{align}   
where $\phi=0$ or $\pi/2$. 

\begin{figure}[!ht]
\begin{center}
	 \includegraphics[width=0.45\textwidth]{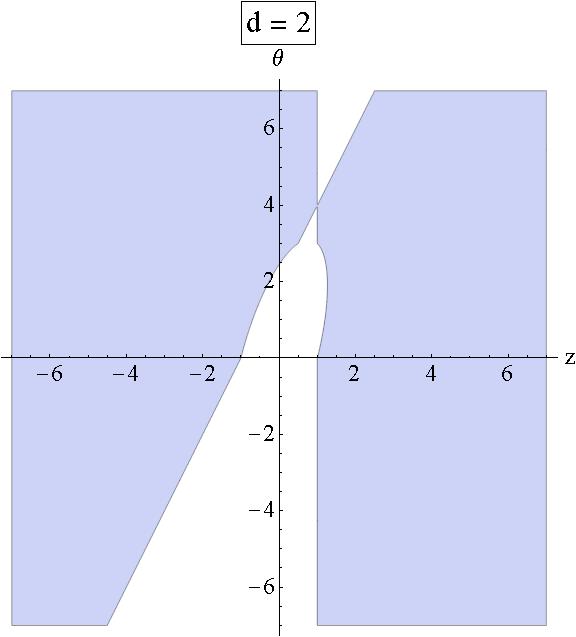} \quad 
	 \includegraphics[width=0.45\textwidth]{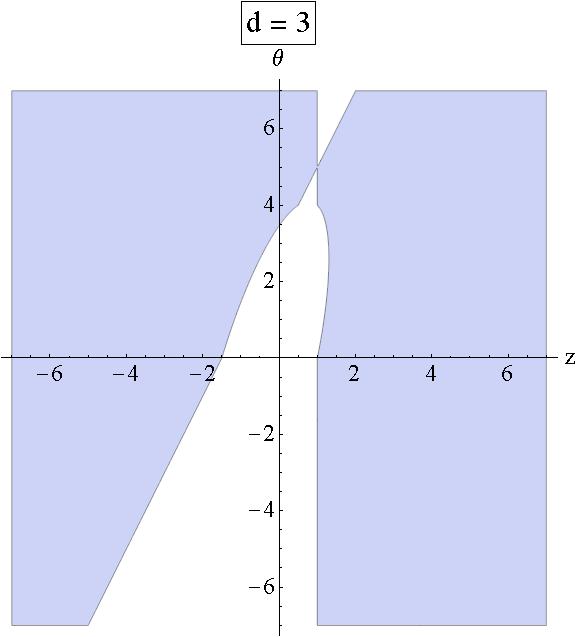}	 
	 \caption{The allowed parameter space of $(z, \theta)$ for the spatial field theory dimensions, $d=2$ and $d=3$, 
	 with $D=d+1$ from the null energy condition. As we increase $d$, the allowed regions around the point 
	 $(z, \theta) =(0, 0)$ are pushed further for negative $z$ and positive $\theta$. 
	 }
	 \label{fig:AllowedRegions}
\end{center}
\end{figure}

From this we can get two independent null energy conditions as 
	\begin{align}   \label{nullECondition}
		& (z-1) (d+2 z)-(1+d) z ~\theta/D + (1+d)~ \theta^2/D^2 \geq 0  \;, \nonumber\\ 
		& (z-1) ( (d+2 z)-(1+d) ~\theta / D ) \geq 0\;,
	\end{align}  
where $d, \theta/D$ and $z$ are the number of spatial coordinates in dual field theory, a parameter associated 
with the hyperscaling violation in the metric (\ref{zeroTMetric}) and dynamical exponent, respectively. 
We see that the null energy condition for Schr\"odinger-type solutions is not particularly simple 
compared to that of the Lifshitz theories. Nonetheless, 
these conditions (\ref{nullECondition}) constrain the allowed values of $(z, \theta)$.
For $z=1$, the first inequality implies that $\theta \le 0$ or $\theta \ge D$, which is 
the same as the null energy condition for $\beta=0$ considered in \S \ref{sec:AdSLC}. And 
this is also similar to the condition obtained in \cite{Dong:2012se}, where both ranges are also realized 
in the string theory constructions. 
How about $z=2$? The conditions become 
	\begin{align}   
		& (d+4)/(d+1) -  2 ~\theta /D + \theta ^2/D^2 \geq 0   \;, \nonumber\\ 
		& (d+4)-(1+d) ~\theta /D \geq 0\;.
	\end{align}  
The first inequality automatically holds and the second condition gives $\frac{\theta}{D} \leq \frac{(d+4)}{d+1}$. 
It seems interesting to find that there is an upper bound for the hyperscaling violation exponent 
for $z>1$, while there is a lower bound for $z<1$. This can be checked in figure \ref{fig:AllowedRegions}. 
For a scale invariant theory, $\theta=0$, both of the conditions become $(z-1)(z +d/2) \geq 0$, 
which gives either the condition $z \geq 1$ or $ z \leq -d/2$.

To get some better understanding, we consider $D=d+1$. This particular choice is made because, in general, 
the spectator coordinate $\xi$ also contributes to our analysis below. 
(Qualitative pictures do not change with different values of $D$.)
Thus we have  
\begin{align}   \label{nullEConditionDdp1}
&  (d+1) (z-1) (d+2 z) - (d+1) z \theta + \theta ^2  \geq 0  \;, \nonumber\\ 
& (z-1)  ( d+2 z- \theta ) \geq 0\;.
\end{align}  
These allowed regions are plotted in figure \ref{fig:AllowedRegions} for $d=2$ and $d=3$ cases, 
which are relevant for the condensed matter systems. 

Let us consider two special cases. For $D=\theta$, we have the two conditions (\ref{nullECondition}) 
to be the same as $(z-1)(2z-1)  \geq 0$, and thus 
	\begin{align}   
 		z>1 ~\quad  \text{or} ~\quad  z<1/2 \;.  
	\end{align}
For $D=d+1, \theta = d$, 
the first one of the conditions (\ref{nullECondition}) is stronger, and thus we have 
	\begin{align}  
			 z> 1/2 + \sqrt{(3d+1)/(d+1)}/2 \quad  \text{or} \quad  z< 1/2 - \sqrt{(3d+1)/(d+1)}/2 \;, 
	\end{align}  
We revisit the null energy condition in the later sections to examine whether 
physically interesting parameter ranges are allowed or not.

\subsection{Propagators from semiclassical approximation}     \label{sec:WKBPropagator}

We consider a scalar field in the background (\ref{zeroTMetric}) and evaluate its geodesic distance traveled by 
a particle with mass $m$ using a semiclassical approximation. In the context of Lifshitz case, the same calculations are 
done in \cite{Dong:2012se}. For the static case, the calculations are the same as the Lifshitz case. 

The differences come into play for the stationary, timelike and general cases. Those cases have an extra conserved 
quantity $\Pi_\xi$ from the $\xi$ direction, which is not fixed by boundary conditions. 
Thus we analytically evaluate the general propagator for some particular values of the parameter $\Pi_\xi$ in this section. 

The action of the particle moving in the background (\ref{zeroTMetric}) is described by 
	\begin{align}     \label{particleAction}
		S = - m \int d \lambda ~ r^{-1 + \theta/D} \sqrt{-r^{-2(z-1)} 
		\left(\frac{dt}{d\lambda}\right)^2 - 2  \frac{dt}{d\lambda} \frac{d\xi}{d\lambda} 
		+ \left(\frac{d r}{d\lambda}\right)^2  + \left(\frac{d x}{d\lambda}\right)^2} \;. 
	\end{align}      
where $\lambda$ is the worldline coordinate. 
We can get a particle geodesic moving along the semiclassical trajectory by extremizing the action (\ref{particleAction}). 
The propagator between two points $x = (t, \xi, x_i)$ and $x'=(t', \xi', x_i')$ on a fixed radius $r = \epsilon$ is 
	\begin{align}    
		G_\epsilon (x',x) \sim \exp \left( S(x',x) \right) \;, 
	\end{align}     
where the coordinates are $(x_f = x', r_f =  \epsilon)$ and $(x_i = x, r_i = \epsilon)$. $r=\epsilon$ is 
the place the dual field theory lives, and $\epsilon$ is related to the scale $R_\theta$. 
The propagator depends only on $\Delta t \equiv t'-t$ and $\Delta x_i \equiv  x_i'- x_i$ for the dual spacetime directions 
due to the space and time translation invariance. It also depends on $\xi$ direction in a special way.

\subsubsection{Static case} 

Let us consider first the static case with $\Delta t=0$. By choosing $\lambda =r$ and 
using $\dot \; \equiv \partial_r$, we get the action 
	\begin{align}    
		S = - m \int dr ~ r^{-1 + \theta/D}\sqrt{1 + \dot x_i^2}\;,
	\end{align}     
where some dimensionful parameters are suppressed.  
The equation of motion is  
	\begin{align}    \label{staticDotxEq} 
		\Pi_x= r^{-1+ \theta/D} \frac{\dot x_i}{\sqrt{1+ \dot x_i^2}} \;, \quad \rightarrow \quad 
		\frac{dx}{dr}= \frac{(r/r_t)^{(D-\theta)/D}}{\sqrt{1-(r/r_t)^{2(D-\theta)/D}}} \;, 
	\end{align}     
where $\Pi_x = r_t^{-2 + 2\theta/D}$ is fixed by the turning point of the geodesic, $dr/dx_i|_{r=r_t}=0$. 
By integrating (\ref{staticDotxEq}) on half of the geodesic motion, we get   
	\begin{align}     
		\frac{|\Delta x_i|}{2} = \frac{\sqrt{\pi} \Gamma\left(\frac{2D-\theta}{2(D-\theta)} \right)}
		{\Gamma \left(\frac{D}{2(D-\theta)} \right)} r_t\,.
	\end{align}     
Thus the total geodesic distance is 
	\begin{align}    \label{staticAction}
		S = -2 m \int_\epsilon^{r_t} dr ~\frac{ r^{-1 + \theta/D}}{\sqrt{1-(r/r_t)^{2(D-\theta)/D}}} 
		= 2 m \frac{D}{\theta} \left(  \epsilon^{\theta/D} - r_t^{\theta/D}  \frac{\sqrt{\pi} 
		\Gamma\left(\frac{2D-\theta}{2(D-\theta)} \right)}{\Gamma \left(\frac{D}{2(D-\theta)} \right)}  \right) \;,
	\end{align}     
where we have neglected higher powers of $\epsilon$. 
In a compact form 
	\begin{align}     
		S = 2m \frac{D}{\theta} \left( \epsilon^{\theta/D}  - \frac{c_{\theta, D}}{2} ~|\Delta x_i|^{\theta/D} \right) \;,
		 \qquad 
		c_{\theta, D} \equiv \left(\frac{2\sqrt{\pi}\,\Gamma\left(\frac{2D-\theta}{2(D-\theta)} \right)}
		{\Gamma \left(\frac{D}{2(D-\theta)} \right)} \right)^{1-\theta/D} \;.
	\end{align}  
Thus, the propagator in the semiclassical approximation is 
	\begin{align}    \label{staticPropagator}
		G(\Delta x_i) \sim  \exp \left[2m \frac{D}{\theta} ~ \epsilon^{\theta/D}\right]  
		 \exp \left[-m \frac{D}{\theta} c_{\theta, D} |\Delta x_i|^{\theta/D}\right] \;.
	\end{align}     
The semiclassical approximation holds for $ m |\Delta x_i|^{\theta/D} \gg 1$ in units of the cross-over scale $R_\theta$. 
In the scale-invariant limit $\theta= 0$, this propagator reduces to the power-law behavior 
	\begin{align}     \label{scalingPropagator}
		G(\Delta x_i) \sim \exp \left[m \log \frac{\epsilon}{|\Delta x_i|} \right] 
		\sim \frac{\epsilon^m}{|\Delta x_i|^m} \;,
	\end{align}     
The propagator (\ref{staticPropagator}) has interesting properties. Compared to the scale invariant limit 
(\ref{scalingPropagator}), the propagator (\ref{staticPropagator}) decay exponentially with the exponent 
$-m |\Delta x_i|^{\theta/D}$, which has the non-trivial $\theta$ dependence. 
These are explained in detail in \cite{Dong:2012se}. Note that this result is only valid for the semiclassical limit.

\subsubsection{Stationary case} 

While it is clear that the static case has the same propagator for the semiclassical approximation 
as the Lifshitz case, there exists an another option for the Schr\"odinger theory :  
a stationary case due to the cross term in the metric (\ref{zeroTMetric}). 
For technical reasons, we consider the parameter ranges $D\geq \theta$ and $z \geq 1$.  

Starting from the action (\ref{particleAction}), 
we would like to consider a stationary case $ \frac{d t}{d\lambda}  + r^{2(z-1)} \frac{d \xi}{d\lambda} =0$. 
Choosing $\lambda =r$ and $\dot \;  \equiv \partial_r $ gives
	\begin{align}  
		S = - m \int dr ~ r^{-1+ \theta/D}\sqrt{r^{2(z-1)} \dot \xi^2 +  1 + \dot x_i^2} \;.
	\end{align}     
The constants of motion are given 
	\begin{align}
		\Pi_\xi = \frac{r^{-1+ \theta/D + 2(z-1)}  \dot \xi}{\sqrt{r^{2(z-1)}  \dot \xi^2 +  1 + \dot x_i^2}} \;, \qquad
		\Pi_i = \frac{r^{-1+ \theta/D }  \dot x_i}{\sqrt{r^{2(z-1)}  \dot \xi^2 +  1 + \dot x_i^2}} \;,
	\end{align}
from which we get $\dot x_i =   \Pi_i / \Pi_\xi   r^{2(z-1)} \dot \xi$. 
Solving these two equations to find 
	\begin{align}     \label{stationaryDotEq}
		&\dot \xi = \frac{\Pi_\xi}{r^{2(z-1)}  \sqrt{r^{-2(1- \theta/D)}  - \Pi_\xi^2 r^{-2(z-1)} - \Pi_i^2} } \;,   \quad
		\dot x_i =   \frac{\Pi_i}{ \sqrt{r^{-2(1-\theta/D)} - \Pi_\xi^2 r^{-2(z-1)} - \Pi_i^2} }  \;.
	\end{align}
The momentum conjugate to $x$ is conserved, and thus the equation of motion can be integrated to 
	\begin{align}      \label{stationaryDotxEq}
		\frac{dx_i}{dr}= \frac{(r/r_t)^{(D-\theta)/D} \left(1- \Pi_\xi^2 ~r_t^{2(D-\theta)/D -2(z-1)} \right)}
		{\sqrt{1-(r/r_t)^{2(D-\theta)/D} - \Pi_\xi^2 ~r^{2(D-\theta)/D -2(z-1)} (1-(r/r_t)^{2(z-1)} )}} \;. 
	\end{align}     
Note that we give special attention to the $\xi$ coordinate and do not impose a boundary condition 
from the expectation that it is an input parameter to define the dual field theory. 
To get (\ref{stationaryDotxEq}), we use 
	\begin{align}
		\Pi_i^2 = r_t^{-2 (D-\theta)/D} - \Pi_\xi^2 ~r_t^{-2(z-1)} \;, 
	\end{align}
from the turning point of the geodesic equation $dx_i/dr |_{r=r_t}=0$.  
For $\Pi_\xi^2=0$, the equation (\ref{stationaryDotxEq}) reduces to the equation (\ref{staticDotxEq}) 
for the static case. 
For $\Pi_\xi^2 = \Pi_{\xi, c}^2 = r_t^{-2 (D-\theta)/D+ 2(z-1)}$, we get vanishing $\Pi_i$ and 
thus $\dot x_i=0$. This means that the massive particle stays at $r=\epsilon$ and does not travel along the 
radial direction. From these analysis, we can conclude that $\Pi_\xi$ behaves as an effective mass 
for the range $0 \leq \Pi_\xi \leq \Pi_{\xi, c}$. 

The action for the stationary case has the following form 
	\begin{align}    \label{stationaryActionA}
		S &= - m \int dr ~ \frac{r^{-2(D-\theta)/D}}{ \sqrt{r^{-2(D-\theta)/D} - \Pi_\xi^2 ~ r^{-2(z-1)} - \Pi_i^2} }  
		\nonumber \\
		&= - m \int dr ~  \frac{r^{-(D-\theta)/D}}{\sqrt{1-(r/r_t)^{2(D-\theta)/D} 
			- \Pi_\xi^2 ~ r^{2(D-\theta)/D -2(z-1)} (1-(r/r_t)^{2(z-1)} )}}  \;.
	\end{align}      
For $\Pi_\xi = 0$, the expression goes back to the static case given in (\ref{staticAction}), and thus the 
massive propagator is the same as the equation (\ref{staticPropagator}).  
One can show that the action (\ref{stationaryActionA}) has an extremum at $\Pi_\xi=0$, 
which is actually a local maximum because its second derivative is less than $0$ for $D>\theta$ and $z>1$. 

For $\Pi_\xi^2 = \Pi_{\xi, c}^2 = r_t^{-2 (D-\theta)/D+ 2(z-1)}$, $\Pi_i =0$ and there is no motion 
along the $x_i$ direction. Thus the action (\ref{stationaryActionA}) becomes 
	\begin{align}      \label{stationaryAction}
		S &= -2 m \int_\epsilon^{r_t} dr ~  \frac{r^{-\frac{D-\theta}{D}}}
			{\sqrt{1 -(\frac{r}{r_t})^{2\frac{D-\theta}{D} -2(z-1)} }} 
		=  2 m \frac{D}{\theta} \left(  \epsilon^{\frac{\theta}{D}} - r_t^{\frac{\theta}{D}}  
			\frac{\sqrt{\pi} \Gamma\left(\frac{(2D-\theta) - 2(z-1)D}{2(D-\theta)-2(z-1)D} \right)}
			{\Gamma \left(\frac{2-z}{2(D-\theta)-2(z-1)D} \right)}  \right)  \;.
	\end{align}  
Even though this action formula (\ref{stationaryAction}) is similar to the static case (\ref{staticAction}), 
there is a crucial difference. Now $r_t$ can be expressed in terms of $\Pi_{\xi, c}$. 
This seems not so illuminating due to the fact that $\Pi_{\xi, c}$ is not associated a definite physical quantity. 

Thus we try to connect this to a physical length scale associated with $\xi$ coordinate. 
We consider that $\xi$ coordinate has the total length, fixed as $L_\xi$, which is a 
defining property of the dual theory for the Schr\"odinger holography. 
We assume that the stationary propagator travels the distance $L_\xi$. 
Integrating the first equation in (\ref{stationaryDotEq}) gives 
	\begin{align}
		L_\xi = r_t^{2-z} \frac{2 }{2-z} c_{L_\xi}  \;, \qquad c_{L_\xi} =
			\frac{\sqrt{\pi}  \Gamma\left(\frac{(2D-\theta) - 2(z-1)D}{2(D-\theta)-2(z-1)D} \right)}
			{\Gamma \left(\frac{2-z}{2(D-\theta)-2(z-1)D} \right)}  \;,
	\end{align}
which gives 
	\begin{align}      \label{stationaryActionLxi}
		S &=2 m \frac{D}{\theta} \left(  \epsilon^{\frac{\theta}{D}} 
			-\hat  c_{L_\xi} L_\xi^{\frac{\theta}{(2-z) D}}  \right)  \;, \qquad 
		 \hat c_{L_\xi} = \left(\frac{2-z}{2} \right)^{\frac{\theta}{(2-z) D}} 
				\left( c_{L_\xi} \right)^{1- \frac{\theta}{(2-z) D}} \;. 
	\end{align}  
Thus the propagator has the form 
	\begin{align}    \label{stationaryLxiPropagator}
		G(L_\xi) \sim  \exp \left[2m \frac{D}{\theta} ~ \epsilon^{\theta/D}\right]  
		 \exp \left[-2 m \frac{D}{\theta} \hat c_{L_\xi} ~L_\xi^{\frac{\theta}{(2-z) D}}\right] \;.
	\end{align}   
This is a reasonable result if we consider the dimension of $L_\xi$, which is $2-z$. 
Note that this case is very special where the geodesic travels only along 
the $\xi$ coordinate. The correlation function can not decay faster than this because there exist the maximum 
distance $L_\xi$ in $\xi$ direction. This is a unique property of the Schr\"odinger type theories. 
  
In summary, one can consider the geodesics along the spatial coordinates and also $\xi$ direction in general. 
For $\Pi_\xi =0$, the stationary geodesic reduces to the static case, while the geodesic only travels 
along $\xi$ direction for $\Pi_\xi =\Pi_{\xi, c}$.  These two extreme cases have different results 
due to the different dimension of $\xi$ coordinate compared to other spatial directions. 
While these propagators reveal different properties for static and stationary cases, we observe that 
the minimal surfaces give the same result in \S \ref{sec:EntanglementEntropy}.

\subsubsection{Timelike}

Until now, we calculated spacelike geodesics. In this section, we would like to perform similar computations 
for the timelike geodesics. With $\lambda =r$ and $\Delta x_i=0$, the action (\ref{particleAction}) gives 
	\begin{align}    
		S = - m \int dr \; r^{-(D-\theta)/D}\sqrt{1 - 2 \dot t \dot \xi - \beta r^{-2(z-1)} \dot t^2}\,.
	\end{align}     
Due to the translation invariance along $t$ and $\xi$, there are corresponding constants of motion  
	\begin{align}
		\Pi_\xi = \frac{r^{-(D-\theta)/D}  \dot t}{\sqrt{1 - 2 \dot t \dot \xi - \beta r^{-2(z-1)} \dot t^2}} \;, \qquad \quad 
		\Pi_t = \frac{r^{-(D-\theta)/D} (\beta r^{-2(z-1)}   \dot t+   \dot \xi )}
			{\sqrt{1 - 2 \dot t \dot \xi - \beta r^{-2(z-1)} \dot t^2}} \;,
	\end{align}
from which we get $\dot \xi = \left(  \Pi_t / \Pi_\xi - \beta r^{-2(z-1)} \right) \dot t$. 
Solving these two equations, we get
	\begin{align}      \label{timelikeDottEq}
		&\dot t = \frac{\Pi_\xi}{\sqrt{2 \Pi_t \Pi_\xi - \beta \Pi_\xi^2 ~r^{-2(z-1)} 
			+  r^{-2\frac{D-\theta}{D}} }} \;,   \quad
		\dot \xi =  \frac{\Pi_\xi \left(  \Pi_t / \Pi_\xi - \beta r^{-2(z-1)} \right) }
			{\sqrt{2 \Pi_t \Pi_\xi - \beta \Pi_\xi^2 ~r^{-2(z-1)} +  r^{-2\frac{D-\theta}{D}} }}  \;.
	\end{align}
Can we impose $\dot \xi =0$ from the second equation? It is only possible when either $\Pi_\xi=0$ or 
$z=1 ~\&~ \Pi_t / \Pi_\xi - \beta=0$, which are trivial cases. 
Thus general timelike geodesics are different from those of the Lifshitz case analyzed in \cite{Dong:2012se}. 

Using the boundary condition for the $t$ coordinate, $dr / dt |_{r=r_t}=0$ at the turning point, 
we fix one of the two constants as
	\begin{align}
		2  \Pi_t \Pi_\xi  = \beta \Pi_\xi^2 ~r_t^{-2(z-1)} - r_t^{-2(D-\theta)/D} \;. 
	\end{align}
Using this we can rewrite (\ref{timelikeDottEq}) as 
	\begin{align}  
		&\dot t = \frac{\Pi_\xi}{\sqrt{r^{-2\frac{D-\theta}{D}} (1-  (r/r_t)^{2\frac{D-\theta}{D}} ) 
			-\Pi_\xi^2 ~\beta r^{-2(z-1)} (1- (r/r_t)^{2(z-1)} ) }} \;,   \nonumber \\
		&\dot \xi =  \frac{\Pi_t   - \Pi_\xi ~ \beta r^{-2(z-1)} }{\sqrt{r^{-2\frac{D-\theta}{D}} 
			(1-  (r/r_t)^{2\frac{D-\theta}{D}} ) -\Pi_\xi^2 ~\beta r^{-2(z-1)} (1- (r/r_t)^{2(z-1)} ) }}  \;.
	\end{align} 
And the action has the following form 
	\begin{align}    \label{timelikeGAction}
		S = - m \int dr ~ \frac{r^{-2(D-\theta)/D}}{\sqrt{r^{-2\frac{D-\theta}{D}} 
			(1-  (r/r_t)^{2\frac{D-\theta}{D}} )  -\Pi_\xi^2 ~\beta r^{-2(z-1)} (1- (r/r_t)^{2(z-1)} )}} \;,
	\end{align}   
where $\Pi_\xi$ is a conserved quantity along the direction $\xi$, on which we don't put physical boundary conditions. 
This brings rather different results on the propagators along the timelike geodesics.  
We choose some special values of $\Pi_\xi$, such as $\Pi_\xi \ll 1$, $\Pi_\xi \gg 1$ and 
$\beta \Pi_\xi^2 = r_t^{-2(D-\theta)/D + 2(z-1)} $, to investigate the system further. 

$\bullet$ For $\Pi_\xi \ll 1$, the geodesic equation for $t$ in (\ref{timelikeDottEq}) reduces to  
	\begin{align}   
		&\frac{dt}{dr} \approx   \Pi_\xi r_t^{\frac{D-\theta}{D}} ~ \frac{(r/r_t)^{\frac{D-\theta}{D}}}
			{\sqrt{1-  (r/r_t)^{2\frac{D-\theta}{D}} } } \;,  \qquad 
		\rightarrow \qquad  \frac{|\Delta t|}{2} \approx \Pi_\xi \sqrt{\pi} r_t^{2-\frac{\theta}{D}} ~ 
			\frac{\Gamma \left(\frac{2D-\theta}{2(D-\theta)}\right)}{\Gamma \left(\frac{D}{2(D-\theta)}\right)}  \;.
	\end{align}
And the action (\ref{timelikeGAction}) is  
	\begin{align}    \label{timelikeActionResultPsmall}
		S &\approx -2 m~\int_\epsilon^{r_t} dr \; \frac{r^{-(D-\theta)/D}}{\sqrt{ 1-  (r/r_t)^{2\frac{D-\theta}{D}} }}  
		= m \frac{2 D}{\theta} \epsilon^{\theta/D} - m \frac{2 D}{\theta} c_{\xi}~ 
			|\Delta t|^{\theta/(2D-\theta)} \;,
	\end{align}     
where    
	\begin{align}   
		&c_{\xi} =  \frac{1}{2 \Pi_\xi } \left(  \frac{ 2 \Pi_\xi  \sqrt{\pi} 
		\Gamma \left(\frac{2D-\theta}{2(D-\theta)}\right)}
		{\Gamma \left(\frac{D}{2(D-\theta)}\right)} \right)^{1- \theta/(2D-\theta)} \;.
	\end{align} 
The corresponding propagator can be obtained by exponentiating this action. 
	\begin{align}    \label{timelikePropagatorSmallP2}
		G(\Delta t) \sim  \exp \left[2m \frac{D}{\theta} ~ \epsilon^{\theta/D}\right]  
		 \exp \left[-2 m \frac{D}{\theta} c_{\xi} ~|\Delta t|^{\frac{\theta}{(2D -\theta)}}\right] \;.
	\end{align}   
This result is similar to the space like geodesics, exponentially decaying with the timelike exponent
$|\Delta t|^{\frac{\theta}{(2D -\theta)}}$ with the $\theta$ dependent power $\frac{\theta}{(2D -\theta)}$.  
Note that this limit is independent of $z$ and is coincident with the case $z=1$. 
This is valid in the regime $ m |\Delta t|^{\frac{\theta}{(2D -\theta)}} \gg 1 $.

$\bullet$ For $\Pi_\xi \gg 1$, one can do the similar analysis.%
\footnote{While we can use Euclidean time $\tau =i t$ along with 
$\eta =- i \xi$ to evaluate the result, we use the Lorentzian time and thus there are $i$'s floating around.   
To get the Euclidean result, we replace $i\sqrt{\beta}$ to $\sqrt{\beta'}$.  }
Geodesic equation for $t$ is  
	\begin{align}    
		&\frac{dt}{dr} \approx   \frac{1}{i\sqrt{\beta}} r_t^{z-1} \frac{(r/r_t)^{z-1}}{\sqrt{ 1- (r/r_t)^{2(z-1)}  }}  \;, 
		 \qquad \rightarrow \qquad  \frac{\Delta t}{2} \approx \frac{1}{i\sqrt{\beta}} \sqrt{\pi} r_t^z ~ 
		 \frac{\Gamma \left(\frac{z}{2(z-1)}\right)}{\Gamma \left(\frac{1}{2(z-1)}\right)} \;.
	\end{align}
And the action is 
	\begin{align}        \label{timelikePiBigAction}
		S  &\approx - \frac{ m}{\Pi_\xi i \sqrt{\beta}} \int dr ~ 
		\frac{r^{z-1-2(D-\theta)/D}}{\sqrt{1- (r/r_t)^{2(z-1)}  }} \nonumber \\
		&=  \frac{ m}{\Pi_\xi i \sqrt{\beta}} \frac{D}{(z-2)D + 2\theta} \epsilon^{\frac{(z-2)D +2\theta}{D}} 
		- \frac{\sqrt{\pi}   m}{2 i \sqrt{\beta} \Pi_\xi (z-1)} \tilde c_{\xi}~ \Delta t^{\frac{(z-2)D+ 2\theta}{Dz}} \;, 
	\end{align}   
where 
	\begin{align}  
		&\tilde c_{\xi} = \frac{\Gamma \left(\frac{(z-2)D+2\theta}{2(z-1)D} \right)}
			{\Gamma \left( \frac{(2z-3)D +2\theta}{2(z-1)D}\right)}
		 \left(  \frac{ i \sqrt{\beta} }{2\sqrt{\pi}} \frac{\Gamma \left(\frac{1}{2(z-1)}\right)}
		 	{\Gamma \left(\frac{z}{2(z-1)}\right)} \right)^{\frac{(z-2)D+2\theta}{D z}} \;.
	\end{align}
Note the differences in the power of the $\Delta t$ compared to the previous case (\ref{timelikeActionResultPsmall}). 
The propagator can be obtained by exponentiating the equation (\ref{timelikePiBigAction}). 

$\bullet$  For $\Pi_t = 0$, one gets an equivalent condition $\beta \Pi_\xi^2 = r_t^{-2(D-\theta)/D + 2(z-1)}$. 
This case is special and $\Pi_\xi$ is determined by the turning point. 
This case turns out to have a similar result to the Lifshitz case \cite{Dong:2012se}. 

To have this restriction $\Pi_t=0$, one needs to impose the condition, $(\beta r^{-2(z-1)}  \dot t+  \dot \xi ) = 0$  
	\begin{align}
		\Pi_\xi = \frac{r^{-(D-\theta)/D}   \dot t}{\sqrt{\beta r^{-2(z-1)} \dot t^2  + 1}}  \;, 
		\quad \rightarrow \quad \dot t=\frac{\Pi_\xi}{\sqrt{r^{-2(D-\theta)/D} - \beta \Pi_\xi^2 r^{-2(z-1)}}} \;.
	\end{align}
Using the fact that the radial derivative vanishes at the turning point along the radial direction,  
we get $\beta \Pi_\xi^2 = r_t^{-2(D-\theta)/D + 2(z-1)} $. Thus there is a correlation between 
$\Pi_\xi$ and the turning point $r_t$, $dr/dt |_{r=r_t}=0$. 
 
The equation of motion can be written
	\begin{align}  
		\frac{dt}{dr}= \frac{r_t^{z-1}}{\sqrt{\beta}} \frac{(r/r_t)^{(D-\theta)/D}}
			{\sqrt{1-(r/r_t)^{2(D-\theta)/D-2(z-1)}}} \;.
	\end{align}     
We can get the expression $\Delta t$ by integrating the equation as 
	\begin{align}     
	\frac{|\Delta t|}{2} &= \frac{ \sqrt{\pi} D~ r_t^z }{\sqrt{\beta} (2D-\theta) } 
	\frac{\Gamma \left( \frac{4D-zD-2\theta}{(2-z) D-\theta}  \right)}
	{\Gamma \left( \frac{6D-3\theta-z D}{2((2-z) D-\theta)}  \right)} \;,
	\end{align}   
which is valid for $2-z-\theta/D >0$. Then the action becomes 
	\begin{align}    
		S  &= -2 m \int_0^{r_t}  dr ~  r_t^{-(D-\theta)/D} \frac{(r/r_t)^{-(D-\theta)/D} }
		{\sqrt{1-(r/r_t)^{2(D-\theta)/D-2(z-1)}}} \nonumber \\
		&=m ~ \frac{2D}{\theta} \epsilon^{\theta/D}-m ~\frac{2D}{\theta} ~\hat c_\xi~ |\Delta t|^{\theta/Dz} \;, 
	\end{align}     
where
	\begin{align}     
		\hat c_\xi =
	 	 \frac{ \sqrt{\pi} \Gamma \left( \frac{2(2-z) D-\theta}{2((2-z) D-\theta)}  \right)}
	 	 {\Gamma \left( \frac{(2-z)D}{2((2-z) D-\theta)}  \right)} 
		\left( \frac{\sqrt{\beta} (2D-\theta)}{2\sqrt{\pi} D} 	
		\frac{\Gamma \left( \frac{6 D-3\theta-z D}{2((2-z) D-\theta)}  \right)}
		{\Gamma \left( \frac{ 4D-zD-2\theta)}{(2-z) D-\theta} \right)} \right)^{\theta/Dz} \;.
	\end{align}     
The propagator for this particular timelike path, with $\Pi_t=0$, is
	\begin{align}       \label{timelikePit=0case}
		G(\Delta t) \sim  \exp\left[2m~\frac{D}{\theta}~ \epsilon^{\theta/D}\right] 
			\exp\left[- 2 m~\frac{D}{\theta}~\hat c_\xi~ |\Delta t|^{\theta/Dz}\right] \;, 
	\end{align}     
in the regime $ m|\Delta t|^{\theta/Dz}\gg 1$. 
This result is very similar to the Lifshitz case \cite{Dong:2012se} with the same power law on $|\Delta t|$ .

\subsubsection{General case}

With the detailed analysis on spacelike and timelike geodesic motions, we try to tackle the general case in this section. 
It turns out that the situation is similar to the timelike case, and we manage to evaluate some particular cases, 
thanks to the special $\xi$ direction present to the Sch\"odinger type metric. 

We consider the action 
	\begin{align}    
		S = - m \int d r \, r^{-(D-\theta)/D}\sqrt{-\beta r^{-2(z-1)} \dot t^2  -2 \dot t \dot \xi + 1 + \dot x_i^2} \;, 
	\end{align}    
where we have set $\lambda =r$.  
There are three constants of motion $\Pi_i, \Pi_t$ and $\Pi_\xi$, which can be solved to give the equations of motion as 
	\begin{align}    
		&\frac{dx}{dr}=  \frac{- \Pi_i}{\sqrt{ 2 \Pi_t   \Pi_\xi - \Pi_i^2   -\beta r^{-2(z-1)} \Pi_\xi^2  + r^{ -2(D-\theta)/D}}} \;,  \nonumber \\
		&\frac{dt}{dr}=   \frac{\Pi_\xi}{\sqrt{ 2 \Pi_t   \Pi_\xi - \Pi_i^2   -\beta r^{-2(z-1)} \Pi_\xi^2  + r^{ -2(D-\theta)/D}}} \;, \\
		&\frac{d\xi}{dr}=  \frac{\Pi_t - \Pi_\xi \beta r^{-2(z-1)}}{\sqrt{ 2 \Pi_t   \Pi_\xi - \Pi_i^2   -\beta r^{-2(z-1)} \Pi_\xi^2  + r^{ -2(D-\theta)/D}}} \;.    \nonumber
	\end{align}    
Also, using the fact that at the turning point $dr/dx |_{r=r_t} =0 , dr/dt |_{r=r_t}=0$, we can derive a relationship 
between $r_t,\Pi_i,\Pi_t$ and $\Pi_\xi$,
	\begin{align}      \label{canstraintGeneralP}
		 2 \Pi_t  \Pi_\xi - \Pi_i^2   -\beta r_t^{-2(z-1)} \Pi_\xi^2 + r_t^{ -2(D-\theta)/D} =0 \;.
	\end{align}    
Plugging this into the geodesic equations, we get  
	\begin{align}     \label{generalDotEq} 
		&\frac{dx}{dr}=  \frac{- \Pi_i}{\sqrt{ -\beta r^{-2(z-1)} (1 - (r/r_t)^{2(z-1)} ) \Pi_\xi^2 + r^{ -2(D-\theta)/D}   (1- (r/r_t)^{ 2(D-\theta)/D})}} \;, \nonumber  \\
		&\frac{dt}{dr}=   \frac{\Pi_\xi}{\sqrt{ -\beta r^{-2(z-1)} (1 - (r/r_t)^{2(z-1)} ) \Pi_\xi^2 + r^{ -2(D-\theta)/D}   (1- (r/r_t)^{ 2(D-\theta)/D})}} \;, \\
		&\frac{d\xi}{dr}=  \frac{\Pi_t - \Pi_\xi \beta r^{-2(z-1)}}{\sqrt{ -\beta r^{-2(z-1)} (1 - (r/r_t)^{2(z-1)} ) \Pi_\xi^2+ r^{ -2(D-\theta)/D}   (1- (r/r_t)^{ 2(D-\theta)/D})}}  \;.  \nonumber
	\end{align}    
	and an expression for the total geodesic distance 
	\begin{align}   
	S = - m \int d r ~ \frac{r^{-2(D-\theta)/D}}{\sqrt{ 2 \Pi_t   \Pi_\xi - \Pi_i^2   -\beta r^{-2(z-1)} \Pi_\xi^2  + r^{ -2(D-\theta)/D}}} \;.
	\end{align}   
as a function of the conserved quantities $\Pi_i, \Pi_t$ and $\Pi_\xi$,  or in terms of the turning point and one 
of these conserved quantities.  
Note that this equation for the general case is slightly more difficult than the timelike geodesic case. 
Thus let us analyze further for some special cases. 

$\bullet$ For $\Pi_\xi \ll 1$,  the equations (\ref{generalDotEq}) reduce to 
	\begin{align}  
		&\frac{dx_i}{dr} \approx   \Pi_i r_t^{\frac{D-\theta}{D}} ~ 
		\frac{(r/r_t)^{\frac{D-\theta}{D}}}{\sqrt{1-  (r/r_t)^{2\frac{D-\theta}{D}} } } \;,  \qquad 
		\rightarrow \qquad  \frac{|\Delta x_i|}{2} \approx \Pi_i \sqrt{\pi} r_t^{2-\frac{\theta}{D}} ~ 
		\frac{\Gamma \left(\frac{2D-\theta}{2(D-\theta)}\right)}{\Gamma \left(\frac{D}{2(D-\theta)}\right)} \;, \\ 
		&\frac{dt}{dr} \approx   \Pi_\xi r_t^{\frac{D-\theta}{D}} ~ 
		\frac{(r/r_t)^{\frac{D-\theta}{D}}}{\sqrt{1-  (r/r_t)^{2\frac{D-\theta}{D}} } } \;,  \qquad 
		\rightarrow \qquad  \frac{|\Delta t|}{2} \approx \Pi_\xi \sqrt{\pi} r_t^{2-\frac{\theta}{D}} ~ 
		\frac{\Gamma \left(\frac{2D-\theta}{2(D-\theta)}\right)}{\Gamma \left(\frac{D}{2(D-\theta)}\right)} \;,
	\end{align} 
and the action 
	\begin{align}    
	S &\approx m \frac{2 D}{\theta} \epsilon^{\theta/D} 
		- m \frac{2 D}{\theta} ~c_\theta~  r_t^{\theta/D}  \;, \qquad  c_\theta =\frac{\sqrt{\pi} 
		\Gamma \left(\frac{2D-\theta}{2(D-\theta)}\right)}{\Gamma \left(\frac{D}{2(D-\theta)}\right)}  \;.
	\end{align}     

To get the expression for $r_t$ in terms of $r_t (|\Delta x_i|, |\Delta t|)$ and $\Pi_t$, we analyze the constraint equation 
$ 2 \Pi_t  \Pi_\xi - \Pi_x^2   + r_t^{ -2(D-\theta)/D} \approx 0$ in the following form 
	\begin{align}    \label{reducedConstraint} 
		& \frac{|\Delta t|}{c_{\theta}} \Pi_t ~ r_t^{-2+\frac{\theta}{D}} - \frac{|\Delta x_i|^2}{4 c_\theta^2}
		 r_t^{-4+2\frac{\theta}{D}}  + r_t^{ -2+ 2\frac{\theta}{D}} \approx 0 \;, 
	\end{align}  
where we leave $\Pi_t$ undetermined because it is related to the motion along $\xi$. 
There are two cases we can consider. When the first term of (\ref{reducedConstraint}) is negligible, 
$r_t$ is proportional to $|\Delta x|$ and we already analyzed this static case. 

When $|\Delta x_i|, |\Delta t| \gg 1$, the last term in (\ref{reducedConstraint}) 
is negligible and this case is more interesting. Then $r_t$ is given by 
	\begin{align}    
	& r_t \approx  \left( \frac{1}{2 c_\theta \Pi_t} \frac{|\Delta x_i|^2}{2 |\Delta t|}\right)^{D/(2D-\theta)} \;,
	\end{align}  
which brings the action in the following form 
	\begin{align}    
	S &\approx m \frac{2 D}{\theta} \epsilon^{\theta/D} 
	- m \frac{2 D}{\theta} c_\theta
	\left( \frac{1}{4 c_\theta \Pi_t}   \frac{|\Delta x_i|^2}{|\Delta t|}\right)^{\theta/(2D-\theta)}  \;.
	\end{align}     
Thus, the propagator for this particular timelike path is given by 
	\begin{align}        \label{timelikePropagatorSmallP}
	G(\Delta t, \Delta x_i) \sim  \exp\left[m \frac{2 D}{\theta}~ \epsilon^{\theta/D}\right] 
	\exp\left[- m \frac{2D}{\theta} c_\theta  
	\left( \frac{1}{4 c_\theta \Pi_t} \frac{|\Delta x_i|^2}{2  |\Delta t|}\right)^{\theta/(2D-\theta)}   \right] \;
	\end{align}     
This propagator is similar to the non-relativistic Schr\"odinger propagator modified by the $\theta$ dependent 
power. Thus the propagator behaves as $G \sim \exp\left[- m \left(\frac{|\Delta x_i|^2}{2  |\Delta t|}\right)^{\theta/(2D-\theta)} \right]$. It is interesting to have this form in the semiclassical 
regime. This would be the generic behavior if $\Pi_t$ and $\Pi_i$ have the same $r_t$ dependence and 
the last term in (\ref{reducedConstraint}) can be ignored.

$\bullet$ For $\Pi_\xi \gg 1$, one has a timelike case. As mentioned above, the propagator can be worked out with 
Euclidean time by replacing $i \sqrt{\beta}$ by $\sqrt{\beta'}$. The equations for $x_i, t$ are 
	\begin{align}   
	 	&\frac{dx_i}{dr}=  \frac{ -\Pi_i /  \Pi_\xi}{\sqrt{ -\beta r^{-2(z-1)} (1 - (r/r_t)^{2(z-1)} ) }} \quad 
	 	\rightarrow  \quad \frac{|\Delta x_i|}{2} =  -\frac{ \Pi_i}{i\sqrt{\beta} \Pi_\xi} 
	 	~r_t^z ~ \frac{\sqrt{\pi} \Gamma \left(\frac{z}{2(z-1)}\right)}{\Gamma \left(\frac{1}{2(z-1)}\right)}  \;, \\ 
		&\frac{dt}{dr}=   \frac{1}{\sqrt{ -\beta r^{-2(z-1)} (1 - (r/r_t)^{2(z-1)} ) }}  \quad 
		 \rightarrow  \quad \frac{|\Delta t|}{2} =  \frac{1}{i\sqrt{\beta} } ~r_t^z ~ 
		 \frac{\sqrt{\pi} \Gamma \left(\frac{z}{2(z-1)}\right)}{\Gamma \left(\frac{1}{2(z-1)}\right)}  \;. 
	\end{align}    
From these equations we conclude $|\Delta x_i| \approx 0$ and $r_t$ is determined only by $|\Delta t|$ for $\Pi_\xi \gg 1$. 
Thus the action and propagator reduce to those of the timelike geodesic case considered in (\ref{timelikePiBigAction}).

\subsection{Holographic energy-momentum tensor}

In holographic approach, background metric is a basic ingredient and  
corresponds to the energy-momentum tensor of the dual field theory. Thus we would like to 
see the effect of $\theta$ on the expectation value of the energy-momentum tensor. 
Lifshitz case is analyzed in \cite{Dong:2012se} and observed that the hyperscaling violation shifts the  
conformal weight of the operator. In this section, we also observe similar effects. 

As mentioned in the introduction, we adapt the effective holographic approach (EHT)\cite{CGKKM}\cite{Dong:2012se} 
and thus consider the metric (\ref{zeroTMetric}) with a radial cutoff at the hypersurface $r=r_c$. 
For the detailed computation, 
we follow the Brown-York procedure \cite{Brown:1992br}\cite{Balasubramanian:1999re} at the hypersurface.%
\footnote{Brown-York procedure is used for the rigorous calculation of the boundary energy-momentum tensor 
in the context AdS in light-cone in \cite{Kim:2010tf}. There the effect of the off-diagonal metric component 
was analyzed to correctly reproduce the total energy and conserved charge by applying \cite{Brown:1992br} 
to the AdS in light-cone. 
}
Usually, it requires to add counter terms to remove diverging pieces for $r_c \rightarrow 0$. 
We do not include the counter terms because our metric (\ref{zeroTMetric}) is valid only upto $r=r_c$ \cite{Dong:2012se}. 

To define appropriate projection on the hypersurface, we need a normal vector for the surface, 
for which we take $n = g_{rr}^{1/2} dr$. The induced metric on the hypersurface is given by 
$\gamma_{\mu\nu} = g_{\mu\nu} - n_\mu n_\nu$, where $\mu, \nu = t, \xi, i, r$. 
The extrinsic curvature and corresponding quasilocal stress tensor are 
	\begin{align}
		&K_{ab} = \gamma_a^\mu \gamma_b^\nu K_{\mu \nu} \;, \qquad K_{\mu\nu} = - n_{\nu ; \mu} \;,  \\
		&\tau_{ab} = K_{ab} - \gamma_{ab} K_{~c}^c \;, \qquad  a, b, c =  t, \xi, i \;.
	\end{align} 
Following the proposal in \cite{Dong:2012se}, we use the prescription to identify the energy-momentum tensor 
at the hypersurface $r=r_c$ as 
	\begin{equation}     \label{QLEPrescription}
		\sqrt{-h} h^{a c} \langle \hat \tau_{c b}\rangle = \sqrt{-\gamma} \gamma^{a c} \tau_{c b} \;,
	\end{equation}
where $h$ is the metric of the dual field theory. $h_{ab}$ is equivalent to $\gamma_{ab}$ upto a conformal factor, 
$h_{ab} = r_c^{2 - 2\theta/D} \gamma_{ab}$.

At the hypersurface $r=r_c$, the induced metric has the form 
	\begin{align}     
		\gamma_{ab} dx^a dx^b =r_c^{-2 + 2\theta/D} \left(-\beta r_c^{-2(z-1)} dt^2 - 2 dt d\xi 
		+ \sum_{i=1}^{d} dx_i^2 \right) \;.
	\end{align}     
Using (\ref{QLEPrescription}), we get 
	\begin{align}     \label{HSET}
		&\langle \hat \tau_{tt}\rangle = -\frac{d + 2 -z -(d+1) \theta/D}{r_c^{d+2-(d+1)\theta/D}} h_{00} \;,  \\ 
		&\langle \hat \tau_{t\xi}\rangle =\langle \hat \tau_{\xi t}\rangle = 
			-\frac{(d+1)(1- \theta/D)}{r_c^{d+2-(d+1)\theta/D}} h_{t\xi} \;,  \\ 	
		&\langle \hat \tau_{ij}\rangle = -\frac{(d+1)(1- \theta/D)}{r_c^{d+2-(d+1)\theta/D}}  h_{ij} \;.
	\end{align}   
From these expressions, we observe that the hyperscaling violation has the effect of modifying the one point function 
of the energy-momentum tensor. The shift is proportional to $\theta$ and reduces to $\theta$ for $D=d+1$. 
Similar results are also observed in correlation functions in \S \ref{sec:correlationFunction}.

\section{Correlation functions}    \label{sec:correlationFunction}

In \S \ref{sec:WKBPropagator}, some basic properties of the propagator are studied in the semiclassical approximation. 
In this section, we study beyond the approximation using the bulk probe scalar in the background (\ref{zeroTMetric}). 
See some of the earlier calculations of the correlation functions in \cite{CorrelationFunctions}.
In \cite{Dong:2012se}, correlation functions are studied in detail for the Lifshitz space with hyperscaling violation. 
There it is reported that the scaling dimension of the scalar operator is changed by $\theta$. 
In particular, it is shifted by the hyperscaling violation exponent for the massless case $m=0$. 

The action for the scalar field is  
	\begin{align}       \label{ScalarAction}
		S = -\frac{1}{2} \int d^{d+3}x~ \sqrt{-g} \left(g^{\mu\nu} 
			\partial_\mu \phi \partial_\nu \phi + m^2 \phi^2 \right)\;.
	\end{align}     
We consider the correlation functions for general $z$ and $\theta=0$ first,  which is important on its own. 
Then we consider $\theta \neq 0$ cases. 
Several different correlators for different vales of $z, \theta$ are considered. 
We generalize the results, which is summarized here (see also the introduction). 
The correlation functions for the Schr\"odinger type theories have the following form 
	\begin{align}    
		\langle\mathcal O(x')\mathcal O(x)\rangle \sim
		\frac{1}{|\Delta t|^{\Delta_+ + (d+1)\theta /2D}} e^{i M \frac{|\Delta \vec x|^2}{2 |\Delta t|} 
		- i \frac{\beta M^2 + m^2}{2M} |\Delta t |}\;, 
	\end{align}     
which is the exact form of the inverse Fourier transform to coordinate space from the following form in the 
momentum space  
	\begin{align}
	 	G(\omega, \vec k, M) \sim  \left( \vec k^2 - 2 M \omega + \beta M^2 + m^2 \right)^{\frac{\Delta_+ - \Delta_-}{2}} \;, 
	\end{align}  
where $\Delta_+, \Delta_-$ are two scaling dimensions associated with the scalar operator $\mathcal O$. 	
The exponential structure $e^{i M \frac{|\Delta \vec x|^2}{2 |\Delta t|}} $ respects the underlying symmetry, 
and the missing dimension between $\vec x$ and $t$ are absorbed by dimensionful parameter $M$ for general $z$.  
The power law behavior $|\Delta t|^{\nu + (d+2)/2}$ is in general depends on the hyperscaling violation exponent $\theta$.

\subsection{Equation of motion}     \label{sec:scalarEOM}

The equation of motion for a scalar field with mass $m$ in the background (\ref{zeroTMetric}) is 
given (in the momentum space) as 
	\begin{align}     \label{scalarFieldEqMom}
		\left(\partial_r^2-\frac{(d+1)(D-\theta)}{D~ r}\partial_r - \vec k^2+ 2 M \omega - \beta \frac{M^2}{ r^{2(z-1)}} 
		-\frac{m^2}{r^{2(D-\theta)/D}}\right)\phi=0 \;,
	\end{align}    
where $\vec k$ and $ \omega $ are Fourier transform of $\vec x$ and $t$, respectively. We treat the $\xi$ direction 
special and replace $\partial_\xi = i M$ for the scalar field \cite{son}. 
Note the last two terms in the equation of motion, which have similar structure and make analysis complicated. 
For zero temperature background, we are concerned for both the boundary behavior at $r\to 0$ 
and the behavior deep in the bulk at $r\to \infty$ of the bulk field. 
For $\beta=0$, things become simpler, and we have some comments in \S \ref{sec:AdSLC}.  

Let us consider the behavior of $\phi$ at the boundary for a small $r$. 
For the parameter range $z>2$ or $\theta<0 $, at least one of the last two terms would have a dominant contribution 
in the equation of motion.%
\footnote{One particular example is the dimensional reduction of D2 brane \cite{Itzhaki:1998dd}, 
which has $z=1$ and $\theta = -1/3$. This was noticed in \cite{Dong:2012se}. 
We thank to Cobi Sonnenschein and Carlos Hoyos for extensive discussions on 
dimensional reduction and related issues for D2 brane.
} 
While these ranges are allowed by the null energy condition and physically interesting, 
it is difficult to analyze. Thus we consider the range $z < 2$ and $\theta>0$.%
\footnote{Note the similar range $z>0 $ and $\theta>0$ for the Lifshitz case \cite{Dong:2012se}. 
The differences come from the fact $g^{tt}=0$ and $g^{\xi\xi} \neq 0$ for the Schr\"odinger-type theories, 
while $g^{tt}\neq 0$ for the Lifshitz theories. 
}
At small $r$, all the terms except the first two are all subdominant. 
Thus we can solve the equation of motion at leading order in $r$ as $\phi\sim r^\nu$,
$\nu=0$ or $\nu = d+2-(d+1)\theta/D$. 
Following the prescription for the correlation functions \cite{sonStarinet}, we impose the regularity condition 
deep in the bulk, $r=\infty$. The full solution in momentum space has the following expansion around 
$r=0$ \cite{Dong:2012se} 
	\begin{align}     \label{CorrelationExpansion}
		\phi = 1 + \cdots + G_k  ~r^{d+2-(d+1)\theta/D} ~( 1 + \cdots )\;,
	\end{align}     
where $\cdots$ refers to the higher order terms in $r$. Due to the normalization, the leading source term 
being normalized to unity, we read off the momentum space correlation function as 
$G_k = G(\vec k, \omega , M, \theta, D, d)$.
We use $k^2 = \vec k^2 - 2 M \omega$, which is an important combination for the Schr\"odinger solution 
in momentum space. 

We also note that, if $M = 0$, the equation of motion is the same as the special case of scalar equation of 
motion in Lifshitz background, namely, $\omega = 0$ there \cite{Dong:2012se}.

\subsection{Schr\"odinger type for general $z$ and $\theta=0$} 

In this section we consider correlation functions of the Schr\"odinger-type theories {\it without} hyperscaling 
violation for its own purposes and for the comparison with the hyperscaling violation cases. 
The equation of motion reduces to 
	\begin{align}     \label{scalarFieldEqMomZerotheta}
		\left(\partial_r^2-\frac{d+1}{ r}\partial_r - k^2  - \beta \frac{M^2}{ r^{2(z-1)}} 
		-\frac{m^2}{r^{2}}\right)\phi=0 \;,
	\end{align}    
Where $k^2 = \vec k^2 - 2 M \omega$. 
Note that, in this case, the term proportional to $m^2$ contribute as the same order as the derivative 
parts, and thus to the scaling dimension of the dual scalar operator. The large $m^2$ limit is corresponding 
to the large scaling dimension limit. 

There are several cases of interest, especially the cases $z=d+1$ and $z=d+2$ for the application of the 
Fermi surface and novel phases as we see in \S \ref{sec:EntanglementEntropy}, 
but these cases require numerical analysis and will not be considered here. 
We would like to review the results for $z=2$ and consider other cases which can be solved exactly.

\subsubsection{Conformal Schr\"odinger for $z=2$}  

This is the most studied case and the system has conformal Schr\"odinger symmetry \cite{son}. 
The analytic results for the correlation functions are available for this case.  
The last two terms in (\ref{scalarFieldEqMomZerotheta}) have the same $r$ dependence. 
The mass term becomes one of the leading contributions, and 
$\Delta = \frac{d+2}{2} \pm \nu=\frac{d+2}{2}\pm\sqrt{\left(\frac{d+2}{2}\right)^2+(\beta M^2 + m^2)}$.
The solution that satisfies the proper boundary condition at $r=\infty$ is
	\begin{align}     \label{z=2Sol}
		\phi= (kr)^{1+d/2} K_{\nu}(kr) \;. 
	\end{align}     
Note that we have normalized the solution at the boundary according to \eqref{CorrelationExpansion}, and thus 
we find the momentum space correlation function as 
	\begin{align}    
		G(k)\sim k^{2\nu} \;,
	\end{align}     
by expanding the modified Bessel function and read off the coefficient as in (\ref{CorrelationExpansion}). 
Fourier transforming back to position space, we find the two-point function to be
	\begin{align}       \label{CFz=2}
		\langle\mathcal O(x')\mathcal O(x)\rangle=\int\frac{d^{d+1}k}{(2\pi)^{d+1}}G(k)e^{ik\cdot(x'-x)}\sim
		\frac{\theta(\Delta t)}{|\Delta t|^\Delta} e^{i M \frac{|\Delta \vec x|^2}{2 |\Delta t|}} \;.
	\end{align}     
Here and below $x'=(t', \vec x')$ and $x=(t, \vec x)$, $\Delta t = t' - t$, $\Delta \vec x = \vec x' - \vec x$ and  
$\mathcal O$ is an operator dual to the massive $\phi$ in the bulk.

\subsubsection{$z=1$}    \label{sec:z=1}

For $z=1$, the scaling dimension of the scalar operator is given by 
$\Delta = \frac{d+2}{2}\pm\sqrt{\left(\frac{d+2}{2}\right)^2 + m^2}$.
The solution for the scalar equation of motion has the same form as (\ref{z=2Sol}) with some 
modification for $\nu = \sqrt{\left(\frac{d+2}{2}\right)^2 + m^2}$. The momentum space 
correlation function is proportional to 
	\begin{align}
		G(k) \sim \left( k^2 + \beta M^2 \right)^{\nu} \;, 
	\end{align}
where we omit the momentum independent factors. 
This specific form can be exactly Fourier transformed back to the coordinate space to find 
	\begin{align}     
		\langle\mathcal O(x')\mathcal O(x)\rangle \sim
		\frac{\theta(\Delta t)}{|\Delta t|^\Delta} e^{i M \frac{|\Delta \vec x|^2}{2 |\Delta t|} 
		- i \frac{\beta M}{2} |\Delta t|}  \;.
	\end{align} 
Thus we observe that the mass $m^2$ contribute to the correlation functions through the scaling dimensions 
of the scalar operator, while the contribution $\beta M^2$ from the term proportional to $\beta$ in (\ref{scalarFieldEqMom}) 
enters in the exponent multiplied by $|\Delta t|$.

\subsubsection{$z=3/2$}

For $z=3/2$, we can solve the differential equation (\ref{scalarFieldEqMomZerotheta}) exactly 
	\begin{align}
		\phi (r) \sim e^{-k r} r^{1+\frac{d}{2}+\nu } U\left( 1/2+ M^2 \beta /(2 k) +\nu , ~1+2 \nu ,~ 2 k r \right) \;, 
	\end{align}
where $U$ represent the confluent Hypergeometric function and $\nu = \sqrt{\left(\frac{d+2}{2}\right)^2 + m^2}$. 
By properly normalize the wave function similar to (\ref{CorrelationExpansion}), 
we can get the momentum space correlation function 
	\begin{align}
		G(k) \sim k^{2 \nu } \frac{ \Gamma \left( 1/2 + M^2 \beta /(2 k) +\nu \right)}
		{\Gamma \left( 1/2 + M^2 \beta /(2k) - \nu \right)} \;,
	\end{align}
where we only keep the energy and momentum dependent part. 
In this case, the Gamma functions are also $k$ dependent in addition to $k^{2\nu}$. 
If $M^2 \beta /(2 k) \ll 1$, the $\Gamma$ functions are independent of $k$. 
Thus, we can Fourier transform back to coordinate space, 
and get the similar result as (\ref{CFz=2}) with modified scaling dimension $\Delta = \frac{d+2}{2} + \nu 
=\frac{d+2}{2} + \sqrt{\left(\frac{d+2}{2}\right)^2+ m^2}$.

\subsection{With hyperscaling violation, $\theta \neq 0$}   \label{sec:corrHSV}

In this section we consider the case with hyperscaling violation and to evaluate 
correlation functions satisfying the differential equation (\ref{scalarFieldEqMom}). 
We would like to see the effects of the hyperscaling violation exponent.

\subsubsection{$m^2 = 0$ case}

The equation of motion in momentum space is 
	\begin{align}     
		\left(\partial_r^2-\frac{(d+1)(D-\theta)}{D~ r}\partial_r - k^2 
		- \beta \frac{M^2}{ r^{2(z-1)}} \right)\phi=0\;.
	\end{align}    
This is the case we can see the effect of the hyperscaling violation clearly. 

For $z=1$, the solution is given by Bessel function with explicit dependence of $\theta$. 
This is similar to the massless case 
considered in \cite{Dong:2012se}. Following the prescription described above, we get an exact result 
	\begin{align}
	G(k) &= c_m  \left(k^2+M^2 \beta \right)^{\frac{1}{2} \left(2+d-\frac{(1+d) \theta }{D}\right)} \;, \quad 
	c_m = 2^{-2-d+\frac{(1+d) \theta }{D}}  \frac{ \Gamma \left(- \frac{2+d}{2}+\frac{(1+d) \theta }{2D}  \right)}{\Gamma \left(\frac{2+d}{2}-\frac{(1+d) \theta }{2 D} \right)} \;,
	\end{align}
upto some numerical factor independent of momentum. 
This function can be exactly evaluated to give the position space correlation function as 
	\begin{align}     \label{corrZeroMassZequal1}
	\langle\mathcal O(x')\mathcal O(x)\rangle &=\int \frac{d^{d}k}{(2\pi)^{d}} \frac{d \omega}{2\pi} 
	 e^{i\vec k\cdot \Delta \vec x } e^{-i\omega \Delta t }  ~ c_m
	\left(k^2+M^2 \beta \right)^{\frac{1}{2} \left(2+d-\frac{(1+d) \theta }{D}\right)}  \nonumber \\ 
	& = \hat c_m ~ 
	\frac{\theta (\Delta t)}{|\Delta t|^{d+2-\frac{(d+1)\theta}{2D}}} 
	e^{i M \frac{|\Delta \vec x |^2}{2 |\Delta t|} - i \frac{\beta M}{2} |\Delta t| }\;,
	\end{align}   	
where $ \Delta t = t' -t $, $\Delta \vec x = \vec x' -\vec x$ and  
	\begin{align}
	\hat c_m = \frac{2^{-\frac{d+4}{2} +\frac{(d+1)\theta}{2D}} M^{d+1-\frac{(d+1)\theta}{2D}} }{\pi i^{d+1-\frac{(d+1)\theta}{2D} + (-1)^d} \Gamma \left(\frac{2+d}{2}-\frac{(1+d) \theta }{2 D} \right) } \;. 
	\end{align}   
Thus we observe two effects. First, the scaling dimension $\Delta_{\theta=0}= d+2$ of the scalar operator is shifted to  
$ \Delta =\Delta_{\theta=0} -\frac{(d+1)\theta}{D}$ due to the hyperscaling violation exponent $\theta$. 
This is similar to the result reported in \cite{Dong:2012se}. Second, the exponent is modified by the time difference factor 
	\begin{align}
		\exp \left(- i \frac{\beta M}{2} |\Delta t| \right) \;,
	\end{align}
which stems from to the modification $k^2 \rightarrow k^2 + \beta M^2$. 
This exponential factor is also present for $\theta=0, z=1$ case. It is rather surprising to find this modification. 
This is an exact result.

At short distance, the two point function is dominated by the large $k$ behavior as 
$G(k)\sim k^{\left(2+d-\frac{(1+d) \theta }{D}\right)}$, and  
the inverse Fourier transform gives 
	\begin{align}      \label{zeroMasszone}
		\langle\mathcal O(x')\mathcal O(x)\rangle 
		\sim \frac{1}{|\Delta t|^{d+2-\frac{(d+1)\theta}{2D}}} e^{i M \frac{|\Delta \vec x|^2}{2 |\Delta t|}}\;.
	\end{align}     
Here we can see the effect of the hyperscaling violation. 
Let us consider the case $\omega = 0$, which corresponds to correlation function for the spacelike separated case 
with $G(k) \sim ( \vec k^2 + \beta M^2 )^{\left(2+d-(1+d) \theta /D\right)/2}$. 
We use saddle point approximation to evaluate this, and we get  
	\begin{align}
		\langle\mathcal O(\vec x')\mathcal O(\vec x)\rangle &
		\sim e^{-\sqrt{\beta} M |\Delta \vec x|}\;,
	\end{align}     
which is valid for $\sqrt{\beta} M |\Delta \vec x| \gg 1$. 
For $\vec k =0$, which is timelike separated case, 
$G(k) \sim (-2M \omega + \beta M^2 )^{\left(2+d-(1+d) \theta /D \right)/2}$. Then again using the 
saddle point approximation, we get 
	\begin{align}
		\langle\mathcal O(t)\mathcal O(t')\rangle 
		\sim e^{-i \frac{\beta M}{2} |\Delta t|}\;.
	\end{align}  
These results do not have the corresponding semiclassical counter part analyzed in \S \ref{sec:WKBPropagator}.

For $z=2$, we have 
\begin{align}     
\left(\partial_r^2-\frac{(d+1)(D-\theta)}{D~ r}\partial_r -k^2 - \beta \frac{M^2}{r^2} \right)\phi=0 \;.
\end{align} 
This equation can be solved explicitly, and the solution is given by Bessel function with explicit dependence of $\theta$. 
In this case we have the conformal dimension of the scalar field as 
\begin{align}      \label{generalScalingDimension}
\Delta = \frac{2+d}{2} -\frac{(d+1)\theta}{2D} \pm \sqrt{ \left(\frac{2+d}{2} 
-\frac{(d+1)\theta}{2D} \right)^2 + M^2 \beta } \;. 
\end{align}
Here we also observe the modification of the scaling dimension compared to that of the $\theta=0$ case.

Following the prescription mentioned above, we get 
\begin{align}
G(k) &=   2^{-\nu} k^{2\nu}\frac{ \Gamma \left(-\nu\right)}{\Gamma \left( \nu \right)} 
\sim  k^{2\nu} \;, \qquad \nu = \sqrt{ \left(\frac{2+d}{2} -\frac{(d+1)\theta}{2D} \right)^2 + M^2 \beta } \;.
\end{align}
Thus the position space correlation function is 
\begin{align}        \label{zeroMassztwo}
&\langle\mathcal O(x')\mathcal O(x)\rangle 
\sim \frac{\theta (\Delta t)}{|\Delta t|^{\frac{d+2}{2} +\nu }} e^{i M \frac{|\Delta \vec x|^2}{2 |\Delta t|}  }
= \frac{\theta (\Delta t)}{|\Delta t|^{\Delta +\frac{(d+1)\theta}{2D} }} e^{i M \frac{|\Delta \vec x|^2}{2 |\Delta t|}  }\;.
\end{align}   
Thus we observe that the naive conformal dimension $\Delta$ is also modified for the correlation function.

\subsubsection{$\theta=D$ case}  

This case is particularly simple and the equation (\ref{scalarFieldEqMom}) only depends on $z$ and $d$. 

For $z=1$, we can evaluate the correlation function explicitly. The equation of motion has the form 
	\begin{align}   
		\left(\partial_r^2 - k^2 - \beta M^2  - m^2  \right)\phi=0\,.
	\end{align} 
Where $k^2 = \vec k^2 - 2 M \omega$. The solution has exponential form 
as $e^{\pm \sqrt{k^2 + \beta M^2  + m^2} r}$. Boundary condition picks up the negative sign and the solution, 
with correct normalization, is 
	\begin{align}
		\phi = e^{- \sqrt{k^2 + \beta M^2  + m^2}~ r} \;.
	\end{align}
The radial expansion factor $d+2-(d+1)\theta/D$ in (\ref{CorrelationExpansion}) becomes unity for $\theta=D$. 
Thus two point function in momentum space is 
	\begin{align}
		G(k) = \sqrt{k^2 + \beta M^2  + m^2} \;,
	\end{align} 
which can be Fourier transform back to position space as 
	\begin{align}
		\langle\mathcal O(x')\mathcal O(x)\rangle 
		= \frac{1}{|\Delta t|^{\frac{d+3}{2}}} e^{i M \frac{|\Delta \vec x|^2}{2 |\Delta t|} 
		- i \frac{\beta M^2 + m^2}{2M} |\Delta t|}\;.
	\end{align}     

Similar to the previous case, we can consider some special cases.  
At short distance, the two point function is dominated by the large $k$ behavior as 
$G(k)\sim k = \sqrt{\vec k^2 - 2M\omega}$, whose inverse Fourier transform gives 
	\begin{align}
		\langle\mathcal O(x')\mathcal O(x)\rangle 
		\sim \frac{1}{|\Delta t|^{\frac{d+3}{2}}} e^{i M \frac{|\Delta \vec x |^2}{2 |\Delta t|}}\;.
	\end{align}     
For $M\omega, M^2 \ll m^2, \vec k^2$, we can use the saddle point approximation to get 
	\begin{align}
		\langle\mathcal O(\vec x')\mathcal O(\vec x)\rangle \sim e^{-m |\Delta \vec x|}\;.
	\end{align}     
This case reduces to (\ref{staticPropagator}) for $D=\theta$. 
For $M^2 \ll m^2$ and $\vec k =0$ which is timelike separated case, we get 
	\begin{align}
		\langle\mathcal O(t')\mathcal O(t)\rangle \sim e^{-i \frac{m^2}{2M} |\Delta t|}\;.
	\end{align}  
These result agrees with the equation (\ref{timelikePropagatorSmallP2}) with some modifications due to 
the presence of $M$ here.

\subsubsection{$z=3/2, \theta=D/2$ case}  

The differential equation has the form 
	\begin{align}    
		\left(\partial_r^2 -\frac{d+1}{2 r} \partial_r  - k^2 - \frac{ \beta M^2 + m^2}{r}  \right)\phi=0\,.
	\end{align} 
In this case analytic solution is available with proper normalization as 
	\begin{align}
		& \frac{\Gamma \left( \frac{5+d}{4}+\frac{ m^2+ M^2 \beta }{2 k}\right)}{ \Gamma \left(\frac{3+d}{2}\right)}
		~ \frac{(2kr)^{3/2+d/2}}{e^{k r}}~ U \left(\frac{5+d}{4} +\frac{m^2+M^2 \beta}{2 k},\frac{5+d}{2},2 k r\right) \;.
	\end{align}
By expanding this solution to $r^{(3+d)/2}$, we can read off 
the two point correlation function at momentum space as
	\begin{align}
		& G(k) =  \frac{(2 k r)^{\frac{3+d}{2}} \Gamma \left(-\frac{d+3}{2}\right) 
		\Gamma \left(\frac{(5+d)}{4} +\frac{ \left(m^2+M^2 \beta \right)}{2 k}\right)}
		{\Gamma \left(\frac{3+d}{2}\right) \Gamma \left(-\frac{1+d}{4}+ \frac{m^2+ M^2 \beta }{2 k}\right)} \;.
	\end{align}
Thus for  $(m^2+ \beta M^2)/k \ll 1$, the momentum two point function has momentum dependence 
as $G(k) \sim k^{\frac{3+d}{2}}$. Back to the position space correlation function, we get  
	\begin{align}
		\langle\mathcal O(x')\mathcal O(x)\rangle 
		\sim \frac{\theta (\Delta t)}{|\Delta t|^{\frac{3d+7}{4}}} e^{i M \frac{|\Delta \vec x|^2}{2 |\Delta t|}}\;.
	\end{align}     
Thus we check again that the scaling dimension of the scalar operator is shifted by $\frac{(d+1)\theta}{2D}$ according to 
the correlation function.

\subsubsection{$\frac{\theta}{D} = \frac{d+1-z}{d+1}$}

Here we would like to consider the case, $\frac{\theta}{D} = \frac{d+1-z}{d+1}$, where 
the logarithmic violation of the area law is observed in \S \ref{sec:novelPhasetheta}. 
In this case the equation of motion does not explicitly depend on $D$ 
	\begin{align}     \label{scalarFieldEqMomLogViolation}
		\left(\partial_r^2-\frac{z}{r}\partial_r -k^2-\beta \frac{M^2}{ r^{2(z-1)}}-\frac{m^2}{r^{2z/(d+1)}}\right)\phi=0 \;,
	\end{align}  
where $ k^2 = \vec k^2 - 2 M \omega $. 

If we consider $m=0$ case, the equation only depends on $z$. 
For $z=2$, the solution reads 
	\begin{align}     
		\phi= (kr)^{(1+z)/2} K_{\nu}(kr) \;, 
	\end{align}
where 
	\begin{align}
		\Delta = \frac{z+1}{2} \pm \nu = \frac{z+1}{2} \pm \frac{\sqrt{(z+1)^2+ 4\beta M^2}}{2} \;.
	\end{align}      
With appropriate normalization, we find the momentum space correlation function as 
	\begin{align}    
		G(k) = 2^{-2\nu} \Gamma (-\nu) / \Gamma (\nu) ~~ k^{2\nu} \;.
	\end{align}     
Fourier transforming back to $(d+1)$-dimensional position space, we find the two-point function to be
	\begin{align}       
		\langle\mathcal O(x')\mathcal O(x)\rangle = 
		\frac{ M^{\nu+d/2}}{i^{\nu +d/2+(-1)^d} \pi 2^{1+\nu} \Gamma (\nu)}
		\frac{\theta(\Delta t)}{|\Delta t|^{\Delta + (d-z+1)/2}} e^{i M \frac{|\Delta \vec x|^2}{2 |\Delta t|}} \;.
	\end{align}     
Thus we explicitly check that the scaling dimension of the operator is shifted 
by $ - \frac{d+1}{2 D} \theta$ including the change in $\nu$, and the correlation function reduces to 
(\ref{CFz=2}) for $z=d+1$, which provides a consistent check. This is similar to Lifshitz case 
\cite{Dong:2012se}, but the magnitude of the shift is reduced by $(d+1)\theta /2 D$. 

Let us briefly mention $z=3/2$ case. The solution of the scalar equation is given by 
	\begin{align}
		\phi \sim e^{-k r} r^{5/2} U \left( \frac{7}{4}+\frac{M^2 \beta }{2k},~\frac{7}{2}, ~2 kr\right)  \;, 
	\end{align}
and the momentum space correlator is 
	\begin{align}
		G(k) \sim  \Gamma \left(\frac{7}{4}+\frac{M^2 \beta }{2 k}\right)/ 
		\Gamma \left(-\frac{3}{4}+\frac{M^2 \beta }{2 k}\right) ~ k^{5/2} \;. 
	\end{align}
Thus, if the momentum dependent part in the Gamma functions can be neglected, $\frac{M^2 \beta }{2 k} \ll 1$, 
one gets  
	\begin{align}       
		\langle\mathcal O(x')\mathcal O(x)\rangle \sim 
		\frac{\theta(\Delta t)}{|\Delta t|^{5/4 + (d+2)/2}} e^{i M \frac{|\Delta \vec x|^2}{2 |\Delta t|}} 
		= \frac{\theta(\Delta t)}{|\Delta t|^{\Delta + (d-z+1)/2}} e^{i M \frac{|\Delta \vec x|^2}{2 |\Delta t|}} \;.
	\end{align}      
Here $\Delta = z+1$. Again the scaling dimension is shifted by $(d-z+1)/2$.  
This happens because the scaling dimension is independent of the spacetime dimensions, while Fourier transform 
has appropriate contributions from them.  

For $z=1$, we encounter a solution 
	\begin{align}
		\phi \sim   r K_{1}(\sqrt{k^2 + \beta M^2} r)  \;, 
	\end{align}
and the momentum space correlator is 
	\begin{align}
		G(k) \sim  ~ (k^2 + \beta M^2)/4 ~ \log \left( (k^2 + \beta M^2)/4 \right)  \;. 
	\end{align}
For $M=0$, which correspond for the large momentum, $G(\vec k) \sim \vec k^2 \log \vec k^2$. 
The position space correlation function is given by $G(\Delta \vec x) \sim 1/|\Delta \vec x|^2$.

There is another case we can evaluate : $d=3, z=2$ and nonvanishing $m$. Equation is 
	\begin{align}     
		\left(\partial_r^2-\frac{2}{r}\partial_r - k^2 - \beta \frac{M^2}{ r^{2}} 
		-\frac{m^2}{r}\right)\phi=0 \;,
	\end{align} 
whose solution gives 
	\begin{align}
		G(k) \sim \frac{ \Gamma \left( \frac{1}{2} \left(1+\frac{m^2}{k}\right)+\nu \right)}
		{ \Gamma \left(\frac{1}{2} \left(1+\frac{m^2}{k}\right)-\nu \right)} ~ k^{2\nu} \;, 
	\end{align}
where $\nu = \sqrt{9/4 + \beta M^2}$. 
For large momentum limit, we get 
	\begin{align}       
		\langle\mathcal O(x')\mathcal O(x)\rangle \sim 
		\frac{\theta(\Delta t)}{|\Delta t|^{\nu + (d+2)/2}} e^{i M \frac{|\Delta \vec x|^2}{2 |\Delta t|}} \;.
	\end{align}  
Thus we also get $\nu + (d+2)/2 = \Delta + (d-1)/2$, and the scaling dimension is shifted 
by $(d-z+1)/2$ compared to $\theta=0$ case.

\subsubsection{$\frac{\theta}{D} = \frac{d+2-z}{d+1}$}
 
For $\frac{\theta}{D} = \frac{d+2-z}{d+1}$, the entanglement entropy is 
proportional to the volume and the area law is extensively violated as discussed in \S \ref{sec:novelPhasetheta}.  
In this case,1 the equation of motion does not explicitly depend on $D$ 
\begin{align}     \label{scalarFieldEqMomVolume}
\left(\partial_r^2-\frac{z-1}{r}\partial_r - k^2 - \beta \frac{M^2}{ r^{2(z-1)}} 
-\frac{m^2}{r^{2(z-1)/(d+1)}}\right)\phi=0 \;,
\end{align}  
where $ k^2 = \vec k^2 - 2 M \omega $. 

If we consider $m=0$ case, the equation only depends on $z$. 
For $z=2$, the solution reads 
	\begin{align}     
		\phi= (kr)^{z/2} K_{\nu}(kr) \;, 
	\end{align}
where 
	\begin{align}
		\Delta = \frac{z}{2} \pm \nu = \frac{z}{2} \pm \frac{\sqrt{(z)^2+ 4\beta M^2}}{2} \;.
	\end{align}      
With appropriate normalization, we find the momentum space correlation function as 
	\begin{align}    
		G(k) = 2^{-2\nu} \Gamma (-\nu) / \Gamma (\nu) ~~ k^{2\nu} \;,
	\end{align}     
Fourier transforming back to $(d+1)$-dimensional position space, we find the two-point function to be
	\begin{align}       
		\langle\mathcal O(x')\mathcal O(x)\rangle = 
		\frac{ M^{\nu+d/2}}{i^{\nu +d/2+(-1)^d} \pi 2^{1+\nu} \Gamma (\nu)}
		\frac{\theta(\Delta t)}{|\Delta t|^{\Delta + (d-z+2)/2}} e^{i M \frac{|\Delta \vec x|^2}{2 |\Delta t|}} \;.
	\end{align}     
Thus we check that the scaling dimension of the operator is shifted by 
$\frac{d+1}{2 D} \theta $ compared to $\Delta_{\theta=0}$. 

Let us briefly mention $z=3/2$ case. The solution of the scalar equation is given by 
	\begin{align}
		\phi \sim e^{-k r} r^{3/2} U \left( \frac{5}{4}+\frac{M^2 \beta }{2k},~\frac{5}{2}, ~2 kr\right)  \;, 
	\end{align}
and the momentum space correlator is 
	\begin{align}
		G(k) \sim \frac{ \Gamma \left(\frac{5}{4}+\frac{M^2 \beta }{2 k}\right)}
		{ \Gamma \left(-\frac{1}{4}+\frac{M^2 \beta }{2 k}\right)} ~ k^{3/2} \;. 
	\end{align}
Thus, if the Gamma functions can be neglected when $\frac{M^2 \beta }{2 k} \ll 1$, 
one gets  
	\begin{align}        \label{exten3halfzCorrelator}
		\langle\mathcal O(x')\mathcal O(x)\rangle \sim 
		\frac{\theta(\Delta t)}{|\Delta t|^{3/4 + (d+2)/2}} e^{i M \frac{|\Delta \vec x|^2}{2 |\Delta t|}} 
		= \frac{\theta(\Delta t)}{|\Delta t|^{\Delta + (d-z+2)/2}} e^{i M \frac{|\Delta \vec x|^2}{2 |\Delta t|}} \;.
	\end{align}      
Again $\Delta = z$, and the scaling dimension is shifted by $\frac{d-z+2}{2} = \frac{d+1}{2 D} \theta$. 
This happens because the scaling dimension depends only on $z$, but the Fourier transform 
has appropriate contributions from the number of spatial dimensions.

\subsection{A scaling argument for $z<2$ and $\theta>0$} 

In general, the scalar equation of motion is difficult to solve analytically, yet we can try to 
guess some general results based on the discussions in this section. 

In the range of the parameters $ z<2$ and $\theta >0$, the last two terms in the differential equation 
 (\ref{scalarFieldEqMom}) 
	\begin{align}     
		\left(\partial_r^2-\frac{(d+1)(D-\theta)}{D~ r}\partial_r -  k^2 - \beta \frac{M^2}{ r^{2(z-1)}} 
	-\frac{m^2}{r^{2(D-\theta)/D}}\right)\phi=0  \;.
	\end{align}     
do not contribute to the scaling dimension of the dual scalar operators. 
Thus the naive scaling dimension is shifted to $\Delta= d+2-(d+1)\theta/D$. 
The equation (\ref{scalarFieldEqMom}) is invariant under the Galilean boost, and the unique combination 
for energy and momenta $k^2 = \vec k^2-2 M \omega$ is maintained. Moreover, there is a scaling symmetry 
in the equation
	\begin{align}     
		r\to\lambda r\,,\quad
		k\to k/\lambda\,,\quad
		M \to M \lambda^{z-2} \;, \quad 
		m\to m/\lambda^{\theta/D}\,,
	\end{align}     
under which the coefficient function $G(k)$ should transform as
	\begin{align}     
		G(k ; M ; m) = \lambda^{\Delta}~ G(k/\lambda ; M \lambda^{z-2} ; m/\lambda^{\theta/D} ) \;. 
	\end{align}     
	
For $z<2$ and $\theta >0$, the momentum space correlation function has the general form  
	\begin{align}
		G(k)\sim k^{\Delta} \cdot F (M/k^{2-z}, ~m/k^{\theta/D}) \;.
	\end{align}
When $M/k^{2-z}, ~m/k^{\theta/D} \ll 1$, $F$ is independent of $k$, and thus we can Fourier transform back to 
the coordinate space to find 
	\begin{align}       
		\langle\mathcal O(x')\mathcal O(x)\rangle \sim 
		\frac{\theta(\Delta t)}{|\Delta t|^{\Delta + (d+1) \theta / (2D)}} e^{i M \frac{|\Delta \vec x|^2}{2 |\Delta t|}} \;.
	\end{align}   
This agrees with several different cases we consider, for example, (\ref{zeroMasszone})(\ref{zeroMassztwo}
(\ref{exten3halfzCorrelator}). 

For the special cases, $z=1$ or $\theta =D$, one of the last two terms in (\ref{scalarFieldEqMom}) lose 
radial dependence and $k$ has some extra contributions. This changes the exponential part of the 
position space correlation function.  
Let us consider $z=1$ and $\theta=D$. Then the coefficient function $G(k)$ should transform as 
the following form and specifically this reduces to a simple function due to $\Delta=1$. Thus 
	\begin{align}     
		G(k ; M ; m) = \lambda^{\Delta}~ G(k/\lambda ; M/ \lambda ; m/\lambda ) \quad \rightarrow 
		\quad \sqrt{k^2 - \beta M^2 - m^2} \;, 
	\end{align}    
whose position space correlation function is 
	\begin{align}
		\langle\mathcal O(x')\mathcal O(x)\rangle 
		= \frac{\theta(\Delta t)}{|\Delta t|^{\frac{d+3}{2}}} e^{i M \frac{|\Delta \vec x|^2}{2 |\Delta t|} 
		- i \frac{\beta M^2 + m^2}{2M} |\Delta t|}\;.
	\end{align}     

For $z=2$ or $\theta=0$, the last two terms in (\ref{scalarFieldEqMom}) contribute to the scaling 
dimension of the corresponding scalar operator. This case is well known conformal Schr\"odinger case. 

Let us conclude this section with some general observations. 
For positive $\theta$, the last term with mass in (\ref{scalarFieldEqMom}) becomes unimportant at short distances. 
Thus the UV behavior of the massive two-point function reduces to the case $m^2=0$ analyzed in \S \ref{sec:corrHSV}, 
which has several distinct results due to the presence of the term proportional to $M^2$ and is 
described in (\ref{corrZeroMassZequal1})(\ref{zeroMassztwo}).  
The long-distance behavior of the massive two-point function is given by the semiclassical approximation 
in \S\ref{sec:WKBPropagator}. 

It will be interesting to investigate the cases with $\theta$ being negative. 
At short distances, the mass term becomes important and there are further technical difficulties to 
analyze the system properly without referring to the UV theory, which is expected to take over at 
the energy scale corresponding 
to $r=r_c$. We have similar technical difficulties even for $\theta=0$ when $z >2$. In this case, at UV, 
the term proportional to $M^2$ dominates over the other terms. It will be interesting to see progress 
for these cases.

\section{Entanglement Entropy}     \label{sec:EntanglementEntropy}

In this section, we would like to consider the entanglement entropy in the context of holographic 
application searching for Fermi surfaces. In \cite{Ogawa:2011bz}, working definition for the system 
with Fermi surfaces was proposed as {\it ``the systems with Fermi surface has the logarithmic violation 
of the area law in their entanglement entropy,"} based on several field theory calculations 
on the theories with fermions \cite{Entanglement}. Lifshitz theories with general dynamical exponent and 
hyperscaling violation exponent were analyzed with null energy condition in 
\cite{Ogawa:2011bz}\cite{Huijse:2011ef}\cite{Dong:2012se}. 
In particular, explicit holographic example is available in \cite{CGKKM}.  
In the context of Lifshitz theories, the logarithmic violation of the area law only exists with hyperscaling violation. 
In this section, we observe that the Schr\"odinger type theories exhibit much richer properties. Surprisingly, 
logarithmic violation of the area law of the entanglement entropy can be also found without hyperscaling violation. 

In \cite{Dong:2012se}, the Lifshitz-type theories with hyperscaling violation is studied more generally, 
and the authors found a range of the exponent $\theta$, which violates the area law, interpolating 
between the logarithmic violation and the volume dependence. In this section, we also observe 
similar properties on the Schr\"odinger-type theories with and without hyperscaling violation.

\subsection{Setup, static and stationary cases} 

From the beginning of the Schr\"odinger holography, it has been clear that the equivalence is formulated as 
codimension 2 correspondence : $(d+1)$-dimensional non-relativistic field theory is equivalent to 
$(d+3)$-dimensional Schr\"odinger background. Thus it is not clear at all that we can uniquely give 
a prescription for the entanglement entropy for this type of theories.%
\footnote{We are grateful to Mukund Rangamani for his valuable comments and discussions on this issue. 
In particular, he gave critical comments that there exists a unique answer for the minimal surface in the context 
of a rotating black hole \cite{Hubeny:2007xt} regardless of static or stationary conditions.  
}
In this section, we would like to make an attempt for clarifying this issue. 

We give careful attentions on the $\xi$ coordinate when we evaluate the minimal surface. 
$\xi$ coordinate has a special role in Schr\"odinger holography and we are supposed to consider a specific 
sector \cite{son}\cite{Kim:2010tf}. 
Let us consider the case when the $\xi$ coordinate spans some finite range, $0 \leq \xi \leq L_\xi$. 
It becomes clear that this subsector provides nothing but the $(2-z)$ dimensional `length scale' 
(or `mass scale' depending on $z$) $L_\xi = \int d\xi$ associated with the $\xi$ direction. 
These can be identified as $L_\xi = \frac{1}{M_\xi}$, where $M_\xi$ is associated a 
`mass scale' associated with the `length scale' $L_\xi$. This is intrinsic from the dual field theory 
point of view due to the fact that the corresponding element in Schr\"odinger algebra commutes with all 
other generators. We consider this as a defining property in the gravity picture. 
Thus we integrate over the entire $\xi$ coordinate 
and treat this as special property when we evaluate the minimal surface.
To describe the dual field theory, we rewrite the metric as 
	\begin{align}    
		ds_{d+3}^2= e^{2A(r)} \left(-\beta e^{2B(r)} dt^2 -2 \left(e^{B(r)} dt \right) \left(e^{-B(r)} d\xi \right) 
		+ dr^2 + \sum_{i=1}^{d} dx_i^2\right)  \;,
	\end{align} 
from which it is clear that the physical length along the $\xi$ direction should be measured by 
$ e^{A(r)-B(r)} d\xi$ rather than $d\xi$. This is clear from the ADM form of the metric 
	\begin{align}    
		ds_{d+3}^2= e^{2A(r)} \left( -\beta \left( e^{B(r)} dt + \beta^{-1} e^{-B(r)} d\xi \right)^2 
		+ \beta^{-1} e^{-2B(r)} d\xi^2 + dr^2 + \sum_{i=1}^{d} dx_i^2\right)  \;,
	\end{align} 
and also give the correct physical dimension to the resulting entanglement entropy.   
This prescription gives the same answer for both the static and stationary cases, 
which are demonstrated explicitly below. See also similar results for the time dependent setup 
in \cite{Hubeny:2007xt}.

\bigskip \bigskip 
\noindent {\it Static case}

For the static case, one choose $|\Delta t| = 0$, and thus effectively the cross term in the metric 
does not contribute to the area. But we still need the integration along the $\xi$ direction. 
From the above explanation, it is clear that the integration should be done for the measure 
$ e^{A(r)-B(r)} d\xi$. This is clearly different from \cite{Dong:2012se}.  

To compute the entanglement entropy, we consider a strip with $\xi$ direction 
	\begin{align}
		0 \le \xi \le L_\xi\;,\quad -l \le x_1 \le l\;,\quad 0 \le x_i \le L\;,\;\quad i = 2,\cdots \;, d
	\end{align}   
in the limit $l \ll L, L_\xi$. 
The strip is located at $r=\epsilon$, and the profile of the surface in the bulk is given by $r=r(x_1)$. 
Thus the area is given by 
	\begin{align} 
		\mathcal A = L^{d-1} L_\xi \int_0^{r_t} dr e^{(d+1) A(r) -B(r) }\sqrt{1+ \left(\frac{dx_1}{dr} \right)^2} \;,
	\end{align}   
where we use $x_1=x_1(r)$ and $dr/dx_1|_{r_t}=0$. 
To signify the physical picture of the dual field theory, we identify the relation $L_\xi = 1/M_\xi$ 
and use both notations.

To obtain the entanglement entropy we extremize $\mathcal A$ and evaluate it on the dominant trajectory.
	\begin{align}   \label{staticEntanglel}
		l = \int_0^{r_t} dr~\frac{e^{-(d+1) \left(A(r) - A(r_t) \right)+  (B(r) -B(r_t))}}
		{\sqrt{1-e^{-2(d+1) \left(A(r) - A(r_t) \right) + 2 (B(r) - B(r_t))}}}\;,
	\end{align}   
and the area is given by 
	\begin{align}     \label{staticEntangleA}
		{\mathcal  A} = L^{d-1} L_\xi \int_\epsilon^{r_t} dr\; \beta^{-1/2} \frac{e^{(d+1) A(r) - B(r) }}
		{\sqrt{1-e^{-2(d+1) \left(A(r) - A(r_t) \right) + 2 (B(r) - B(r_t))}}}\;.
	\end{align}   
Thus the entanglement entropy for a strip in the stationary case of the general metric (\ref{ADMzeroTMetric}) is given by 
	\begin{align}    \label{staticEntangleE}
		\mathcal  S = \frac{M_{Pl}^{d+1}}{4} \mathcal  A
	\end{align}   
with $M_{Pl}$ the $(d+3)$-dimensional Planck constant.

\bigskip \bigskip
\noindent {\it Stationary case} 

For the stationary case, we consider the ADM form of the metric (\ref{ADMzeroTMetric}) with the condition 
	\begin{align}
		e^{B(r)} dt + \beta^{-1}  e^{-B(r)} d\xi =0 \;.
	\end{align}
Again, we compute the entanglement entropy for a strip times the direction $\xi$ as 
	\begin{align}
		0 \le  \xi \le L_\xi \;,\quad -l \le x_1 \le l\;,\quad 0 \le x_i \le L\;,\quad i = 2, \cdots \;, d
	\end{align}   
with the assumption, $l \ll L, L_\xi$. Similarly, the expression for the surface is 
	\begin{align} 
		\mathcal A &= L^{d-1} L_\xi \int_0^{r_t} dr  \beta^{-1/2} e^{(d+1)A(r) -B(r) }
		\sqrt{1+ \left(\frac{dx_1}{dr} \right)^2}  \;.
	\end{align}  
Thus we quickly realize that the entanglement entropy for the stationary case is the same as 
the static case. The result is given by the equations, (\ref{staticEntanglel}), (\ref{staticEntangleA})
and (\ref{staticEntangleE}).

\subsection{General Entanglement regions}      \label{sec:generalEntangleRegion}

While we perform the calculation of the entanglement entropy for the strip geometry, 
it can be shown to hold for more general surfaces. 
As we already see, static and stationary cases give the same answer. 
 
Let us consider a general surface 
	\begin{align}   
		x_d = \sigma(x_i, \xi)\;, \qquad i =1, \cdots \;, d-1 \;, 
	\end{align}   
at $r=\epsilon$, where the dual field theory lives. 
Then the surface that extremizes the area will be 
	\begin{align}   
		x_d = \Sigma(x_i, \xi, r)\;,\;\Sigma(x_i, \xi, 0) = \sigma(x_i, \xi)\,.
	\end{align}   
The pullback of the metric onto $\Sigma$ has the induced metric 
	\begin{align}   
		ds_\Sigma^2 &=e^{2A}\left(\left[ 1 + (\partial_r \Sigma)^2\right] dr^2 + 2 
		 \partial_r \Sigma \partial_i \Sigma\, dr dx_i +  \left(\delta_{ij}+ \partial_i \Sigma 
		 \partial_j \Sigma \right) dx^i dx^j \right. \nonumber \\
		& \left.  \qquad 
		+ \left( \beta^{-1} e^{-2B(r)} + ( \partial_\xi  \Sigma)^2 \right)  d\xi^2 
		+ 2 \partial_\xi \Sigma  ( \partial_r \Sigma dr + \partial_i \Sigma dx^i ) d\xi \right) \;, 
	\end{align}   
and the area reads
	\begin{align}   \label{generalEntangleArea}
		\mathcal  A = \int d^{d-1} x \,dr \; d \xi \,e^{(d+1)A(r)} \sqrt{(\partial_\xi \Sigma)^2 
		+ \beta^{-1} e^{-2B(r)}  (1 + (\partial_r \Sigma )^2 + (\partial_i \Sigma)^2 ) }\,.
	\end{align}   
The equation of motion gives the conserved currents
	\begin{align}   
		J_\xi &= \frac{e^{(d+1)A(r)} \partial_\xi \Sigma}{\sqrt{(\partial_\xi \Sigma)^2 + \beta^{-1} e^{-2B(r)}  
		(1 + (\partial_r \Sigma )^2 + (\partial_i \Sigma)^2 ) }} \;, \nonumber\\
		J_r &= \frac{\beta^{-1} e^{(d+1)A(r)} e^{-2B(r)} \partial_r \Sigma}{\sqrt{(\partial_\xi \Sigma)^2 
		+ \beta^{-1} e^{-2B(r)}  (1 + (\partial_r \Sigma )^2 + (\partial_i \Sigma)^2 ) }} \;, \nonumber \\
		J_i&= \frac{\beta^{-1} e^{(d+1)A(r)} e^{-2B(r)} \partial_i \Sigma}{\sqrt{(\partial_\xi \Sigma)^2 
		+ \beta^{-1} e^{-2B(r)}  (1 + (\partial_r \Sigma )^2 + (\partial_i \Sigma)^2 ) }}\;.
	\end{align}   
While these equations are expressed with equal footing for $J_r, J_i$ and $J_\xi$, $J_\xi$ plays a special role. 
Thus we consider $J_\xi$ as an input parameter. 
One can express $J_r$ and $J_i$ in terms of $J_\xi$ as 
	\begin{align}
		&J_r = J_\xi \beta^{-1} e^{-2B(r)} \partial_r \Sigma /  \partial_\xi \Sigma \;, \qquad 
		J_i = J_\xi \beta^{-1} e^{-2B(r)} \partial_i \Sigma /  \partial_\xi \Sigma  \;. 
	\end{align}
These constants of motion enables us to solve the equations of motion 
	\begin{align}   
		\partial_\xi \Sigma &= \frac{ J_\xi \beta^{-1/2} e^{-B(r)}}{\sqrt{ e^{2(d+1)A(r)} - J_\xi^2 - \beta e^{2B(r)} 
		(J_r^2 + J_i^2)   }}  \;, \nonumber\\
		\partial_r \Sigma &= \frac{ J_r \beta^{1/2} e^{B(r)}}{\sqrt{ e^{2(d+1)A(r)} - J_\xi^2 - \beta e^{2B(r)} 
		(J_r^2 + J_i^2)   }}  \;,  \nonumber\\
		\partial_i \Sigma &= \frac{ J_i \beta^{1/2} e^{B(r)}}{\sqrt{ e^{2(d+1)A(r)} - J_\xi^2 - \beta e^{2B(r)} 
		(J_r^2 + J_i^2)   }}  \;.
	\end{align}   
The combined value of $J_r^2 + J_i^2$ is determined in terms of the input $J_\xi$ and the turning point $r_t$ as 
$J_r^2 + J_i^2 = \beta^{-1} e^{-2 B(r_t)} \left( e^{2(d+1)A(r_t)} - J_\xi^2 \right)$.
Then the area (\ref{generalEntangleArea}) is given by 
	\begin{align}   
		\mathcal  A &= A_{\xi, x_i}   \int dr ~  \frac{e^{2(d+1)A(r)} \beta^{-1/2} e^{-B(r)} }{ \sqrt{ e^{2(d+1)A(r)} 
		(1 - e^{-2(d+1)(A(r)-A(r_t))+ 2(B(r)-B(r_t))} ) - J_\xi^2  (1-e^{2B(r)-2B(r_t)}  )  }} \;. 
	\end{align}   
Where we used that the area function depends only on $r$, and thus the integrals over the spaces $\xi$ and $x_i$ 
can be trivially done to give $A_{\xi, x_i}$. Note that the area is function of a constant of motion $J_\xi$, which 
we don't expect to fix by the boundary condition. 
To extremize the area, we set $J_\xi =0$. Then the result is reduced to the strip geometry (\ref{staticEntangleA}). 
This analysis signals that to extremize the minimal area, the surface $\Sigma$ does not have explicit $\xi$ dependence. 
Thus we show that the explicit calculation with strip geometry holds in more general setup.

\subsection{Entanglement entropy for general $z$ and $\theta=0$} 

In this section we present the result of the entanglement entropy for the Schr\"odinger-type theories 
{\it without} hyperscaling violation, $\theta=0$. As far as we are aware, the holographic calculation 
for the entanglement entropy for the Schr\"odinger-type theory was not explicitly done before due to the 
conceptual difficulties explained above.

Both for the static and the stationary cases, 
we need $e^{2B} =  r^{-2(z-1)}$ in the metric (\ref{zeroTMetric}) in addition to 
the warp factor $e^{2A} = R^2/r^2$, where $R$ is the curvature radius. 
Thus we have $e^{-(d+1)A + B} = R^{-(d+1)} r^{d-z+2}$. 

From (\ref{staticEntanglel}) and (\ref{staticEntangleA}), we get  
	\begin{align}     \label{staticEntanglelSchr}
		l &= \int_0^{r_t} dr ~\frac{(r/r_t)^{d-z+2}}{\sqrt{1- (r/r_t)^{2(d-z+2)}}} 
		= \sqrt{\pi} r_t \frac{\Gamma \left( \frac{d-z+3}{2(d-z+2)} \right)}{\Gamma \left( \frac{1}{2(d-z+2)} \right)} \;,
	\end{align}   
and 
	\begin{align}   
		{\mathcal  A}       \label{staticEntangleASchr}
		& = L^{d-1} L_\xi \beta^{-1/2} R^{d+1} \int_\epsilon^{r_t} dr~  
		\frac{ r^{-d+z-2}}{\sqrt{1-(r/r_t)^{2(d-z+2)}}}   \nonumber \\
		&= \frac{\beta^{-1/2} R^{d+1} }{ (d-z+1)} \left(\frac{L^{d-1} L_\xi }{\epsilon^{d-z+1}} 
		- \frac{L^{d-1} L_\xi}{r_t^{d-z+1}} \frac{\sqrt{\pi} \Gamma \left( \frac{d-z+3}{2(d-z+2)} \right)}
		{\Gamma \left( \frac{1}{2(d-z+2)} \right)}   \right) \;.
	\end{align}   
Thus the entanglement entropy for a strip in the general metric (\ref{zeroTMetric}) is 
	\begin{align}    \label{entanglementEntropySchr}
		\mathcal  S  
		&= \frac{(R M_{Pl})^{(d+1)}}{4  (d-z+1)}  \left( \left(\frac{L}{\epsilon} \right)^{d-1}  
		\left(\frac{L_\xi}{\epsilon^{2-z}} \right)
		  -  c_z ~  \left(\frac{L}{l} \right)^{d-1}  \left(\frac{L_\xi}{l^{2-z}} \right)  \right) \;,
	\end{align}   
where 
	\begin{align}
		c_z  = \left(  \frac{\sqrt{\pi}  \Gamma \left( \frac{d-z+3}{2(d-z+2)} \right)}
		{\Gamma \left( \frac{1}{2(d-z+2)} \right)}    \right)^{d-z+2}
	\end{align}
and with $M_{Pl}$ the $(d+3)$-dimensional Planck constant. 
This is the general expression for the entanglement entropy for the Schr\"odinger-type theories 
with general dynamical exponent $z$. Note that the dimensionless combinations of the terms related to $L_\xi$.

It is interesting to see the case $z=2$ explicitly. 
	\begin{align}    \label{entanglementEntropySchrz=2}
		\mathcal  S_{z=2}  
			&= \frac{(R M_{Pl})^{(d+1)}}{4  (d-1) M_\xi }  ~  \left( \left(\frac{L}{\epsilon} \right)^{d-1}
		  -  c_{z=2} ~  \left(\frac{L}{l} \right)^{d-1}  \right) \;.
	\end{align}   
This is a particularly interesting result. For $z=2$, the theory is known to have codimension 2 holographic 
equivalence. This formula manifestly reveals that the area law is actually hold for 
$(d+1)$-dimensional boundary theory. We use the notation $L_\xi = \frac{1}{M_\xi}$, 
where $\xi$ becomes dimensionless for $z=2$. 
For $z=1$ , one gets 
	\begin{align}   
		\mathcal  S_{z=1}  
		&= \frac{(R M_{Pl})^{(d+1)}}{4 d}  \left(\left( \frac{L  }{\epsilon} \right)^{d-1} 
		\left(\frac{L_\xi}{\epsilon} \right)
		- c_{z=1}~ \left( \frac{ L}{l}\right)^{d-1}  \left(\frac{L_\xi}{l} \right)  \right)  \;.
	\end{align}   
Thus the entanglement entropy for $z=1$ is similar to that of the Lifshitz-like theories for $d+1$ dimensions.

\subsubsection{Novel phases with $d+1 < z < d+2$ }  \label{sec:novelPhasez}

Furthermore, note that this case has far more interesting behavior in view of searching Fermi surface holographically. 
Even though the hyperscaling violation term is not present as we take $\theta=0$, 
it is possible to have logarithmic violation of area law for $z=d+1$, which is surprising. 
This can be explicitly checked from equations (\ref{staticEntanglelSchr}) and (\ref{staticEntangleASchr}).
 
For $z=d+1$, the area gives logarithmic dependence of $l$ instead of the 
power law, and the entropy reads 
	\begin{align}       \label{logEntropyz}
		\mathcal  S_{z=d+1}  
		&= \frac{(R M_{Pl})^{(d+1)}}{4 \beta^{1/2}} \left( \frac{L^{d-1} }{M_\xi} \right) 
		\log \left(\frac{2l}{\epsilon} \right)  \;.
	\end{align}   
This shows a logarithmic violation of the area law even {\it without} hyperscaling violation and 
signals the presence of a Fermi surface in the dual field theory, according to \cite{Ogawa:2011bz} . 
This behavior is distinct from the that of the Lifshitz-type theories with only diagonal components 
in the metric. It will be interesting to study the properties of the Schr\"odinger background in detail 
for $z=d+1$. 

For $z=d+2$, the combined warp factor becomes $e^{(d+1)A - B} =  R^{d+1}$, and 
thus the area (\ref{staticEntangleA}) turns into  
	\begin{align}   
		\mathcal  S_{z=d+2}  
		&= \frac{(R M_{Pl})^{(d+1)}}{2 \beta^{1/2}} \left(\frac{L^{d-1}  ~l}{M_\xi} \right) \;.
	\end{align}     
This entanglement entropy has an extensive contribution, proportional to the volume of the surface at $r=\epsilon$. 
This happens because the surface containing the entanglement region does not have any profile along the $r$ direction, 
and thus the entropy is proportional to the volume of the region. Those features are explained in \cite{Dong:2012se} 
in the context of Lifshitz-type theories with hyperscaling violation. 

From these observations, we conclude that there exist new types of phases of matter for  
	\begin{align}      \label{novelzRange}
		d+1 < z < d+2 \;, 
	\end{align}
At $z=d+1$, the Schr\"odinger-type theories develop a logarithmic violation of area law, which signals 
the presence of Fermi surface, while the area law is extensively violated at $z=d+2$.

\subsection{Entanglement entropy with hyperscaling violation} 

In this section, we would like to generalize the discussion of Entanglement entropy for the case with 
hyperscaling violation, $\theta \neq 0$. For this general case, we can identify $e^{2A} =  r^{-2(D-\theta)/D}$ 
and $e^{2B} =  r^{-2(z-1)}$ from the metric (\ref{zeroTMetric}). Thus we have 
$e^{-(d+1)A + B} =  r^{(d+1) (D-\theta)/D -z+1} \equiv r^{\alpha}$, where $\alpha = (d+1) (D-\theta)/D -z+1$. 

Following the similar steps, we can evaluate the expressions (\ref{staticEntanglel}) and (\ref{staticEntangleA}) 
	\begin{align}   
		l &= \int_0^{r_t} dr~ \frac{(r/r_t)^{\alpha}}{\sqrt{1- (r/r_t)^{2\alpha}}} 
		= \sqrt{\pi} r_t \frac{\Gamma \left( \frac{1+\alpha}{2\alpha} \right)}{\Gamma \left( \frac{1}{2\alpha} \right)} \;, 
	\end{align}   
and 
	\begin{align}  
		{\mathcal  A} 
		&= L^{d-1} L_\xi \int_\epsilon^{r_t} dr~ \frac{ \beta^{-1/2} r^{-\alpha}}{\sqrt{1-(r/r_t)^{2\alpha}}} 
		= \frac{\beta^{-1/2} L^{d-1} L_\xi }{ (\alpha-1)} \left( \frac{1}{\epsilon^{\alpha-1}} 
		- \frac{c_\theta}{l^{\alpha-1}} \right)	\;, 
	\end{align}   
where 
	\begin{align}
		\alpha = d-z+2 -\frac{(d+1)\theta}{D} \;, \qquad 
		c_{\theta} = \left( \frac{\sqrt{\pi} \Gamma \left( \frac{1+\alpha}{2\alpha} \right)}
		{ \Gamma \left( \frac{1}{2\alpha} \right) } \right)^\alpha \;.
	\end{align} 
The entanglement entropy for a strip in the general metric (\ref{zeroTMetric}) is 
	\begin{align}      \label{entanglementEntropyHyper}
		\mathcal  S 
		&= \frac{(R M_{Pl})^{(d+1)} }{4 (\alpha-1)} 
		\left(  \left(\frac{\epsilon}{R_\theta}\right)^{(d+1)\theta/D} \frac{L^{d-1} L_\xi }{\epsilon^{d-z+1}} 
		-  c_{\theta} ~\left(\frac{l}{R_\theta}\right)^{(d+1)\theta/D} \frac{L^{d-1} L_\xi }{l^{d-z+1}}  \right) \;,
	\end{align}   
where $R_\theta$ is a scale in which the hyperscaling violation becomes important. 
This result is the generalization of the entanglement entropy (\ref{entanglementEntropySchr}) by including the hyperscaling 
violation $\theta \neq 0$. For $\theta=0$, the result is reduced to (\ref{entanglementEntropySchr}). We observe further 
modification of the entropy by an additional power of (length)$^{-(d+1)\theta/D}$. This effect comes from the fact that 
the metric has dimension $\theta/D$ and the entanglement entropy for the $(d+1)$-dimensional region would get the contribution.    

For $\frac{\theta}{D} = \frac{2-z}{d+1}$, the entanglement entropy (\ref{entanglementEntropyHyper}) reduces to 
	\begin{align}    
		\mathcal  S 
		&= \frac{(R M_{Pl})^{(d+1)} }{4 (d-1) } \left( \frac{L_\xi}{R_\theta^{2-z}} \right) 
		\left( \frac{L^{d-1} }{\epsilon^{d-1}} -c_{\theta} ~\frac{L^{d-1} }{l^{d-1}}  \right) \;, \qquad 
		c_{\theta} = \left( \frac{\sqrt{\pi} \Gamma \left( \frac{1 +d }{2 d} \right)}
		{ \Gamma \left( \frac{1}{2 d} \right) } \right)^d \;, 
	\end{align}   
where the result shows clear area law dependence.

\subsubsection{Novel phases with $\frac{d+1-z}{d+1}  < \frac{\theta}{D} < \frac{d+2-z}{d+1} $ }  \label{sec:novelPhasetheta}

Following the similar discussion in \S \ref{sec:novelPhasez}, 
we consider some novel phases of the dual field theory with the hyperscaling violation. 
For $\frac{\theta}{D} = \frac{d+1-z}{d+1} $, the minimal surface gives logarithmic dependence of $l$, and the entropy reads 
	\begin{align}     \label{logEntropytheta}
		\mathcal  S_{\theta =\frac{d+1-z}{d+1} D}  
		&= \frac{(R M_{Pl})^{(d+1)}}{4 \beta^{1/2}} \left(\frac{L^{d-1}}{M_\xi} \right) 
		\log \left(\frac{2l}{\epsilon} \right)  \;.
	\end{align}   
This shows a logarithmic violation of the area law with hyperscaling violation, which 
signals the presence of a Fermi surface in the dual field theory, according to \cite{Ogawa:2011bz} . 
The expressions are the same as (\ref{logEntropyz}), as it is clear from the mathematical expressions.

\begin{figure}[!ht]
\begin{center}
	 \includegraphics[width=0.45\textwidth]{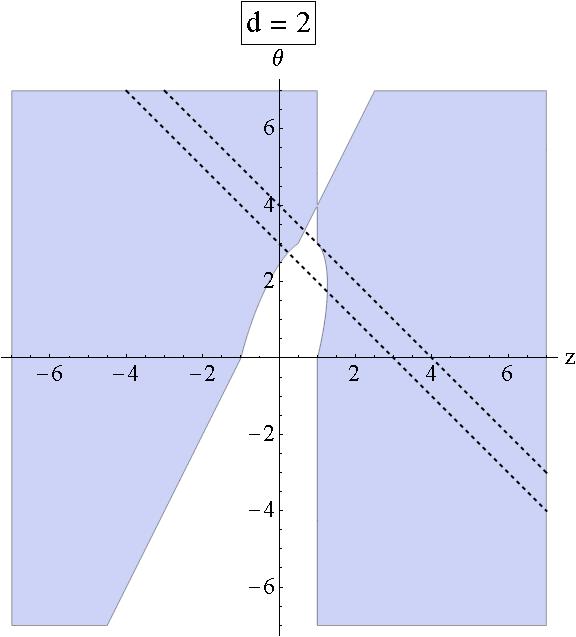} \quad 
	 \includegraphics[width=0.45\textwidth]{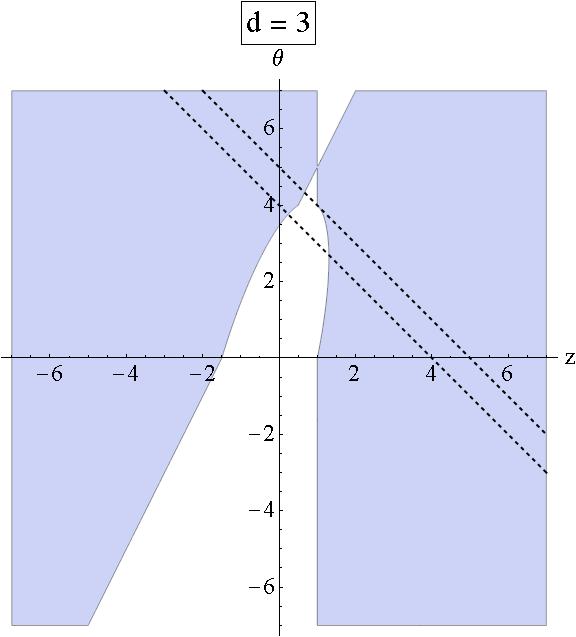}	 
	 \caption{The parameter ranges of $(z, \theta)$ for the novel phases are plotted for $d=2$ and $d=3$. 
	 The plot assumes $D=d+1$. The novel phases lie in the region between the black dashed lines. 
	 The blue background is allowed regions from the null energy condition.  
	 }
	 \label{fig:NovelPhaseAllowedRegions}
\end{center}
\end{figure}

For $\frac{\theta}{D} = \frac{d+2-z}{d+1} $, $e^{(d+1)A - B} =  R^{d+1} R_\theta^{z-2-d}$, and 
thus the minimal surface area (\ref{staticEntangleA}) becomes  
	\begin{align}    
		\mathcal  S_{\theta = \frac{d+2-z}{d+1} D}  
			&= \frac{(R M_{Pl})^{(d+1)}}{2 \beta^{1/2}} \left(\frac{L^{d-1} ~l}{R_\theta^{d+2-z} M_\xi} \right)  \;.
	\end{align}     
This entanglement entropy has an extensive contribution, proportional to the volume of the surface at $r=\epsilon$. 
Thus in the following parameter range, 
	\begin{align}       \label{novelthetaRange}
		\frac{d+1-z}{d+1} < \frac{\theta}{D} < \frac{d+2-z}{d+1} \;, 
	\end{align}
the Schr\"odinger-type theories develop new violations of area law, and thus some novel phases, similar to the 
cases analyzed in \S \ref{sec:novelPhasez}. This parameter regions are plotted in figure \ref{fig:NovelPhaseAllowedRegions}.  
It is interesting to observe that for $\theta=0$, the range of the parameters in (\ref{novelthetaRange}) reproduce the 
previous result in (\ref{novelzRange}).  

For $z=2$ and $D=d+1$, the range (\ref{novelthetaRange}) reduces to 
	\begin{align}       	
		d-1 < \theta < d \;, 
	\end{align}
which reproduces the same range of the novel phases observed in \cite{Dong:2012se}.  

It will be interesting to investigate the theories in the parameter ranges. We hope to report the properties 
of these theories soon.

\section{String theory realizations}    \label{sec:stringConst}

In this paper we took the effective holographic approach (EHT) by constructing our metric 
with appropriate symmetries, valid for certain range of energy scale.  
It will be important to have a direct construction of the Schr\"odinger metric with hyperscaling violation 
from string theory, which will provide the UV completion. In the literature, there already exist several papers 
constructing Schr\"odinger type metric \cite{topdownSchrMetric}\cite{topdownSchrMetric2} for several values of $z$ 
using null Melvin twist \cite{nullMelvinTwist}. See also \cite{SchrSolutions} for the Schr\"odinger solutions 
with general dynamical exponent $z$. In \cite{topdownSchrMetric}, many different solutions and 
their finite temperature generalizations are constructed.%
\footnote{We thank to Yaron Oz for the discussions related to the non-relativistic branes.} 
We concentrate on the zero temperature null Melvin 
twist of D$p$ brane solutions, whose dimensional reductions are expected to provide the Schr\"odinger metric 
with hyperscaling violation. 

The metric \cite{topdownSchrMetric} is given by 
	\begin{align}     \label{nonrelBrane}
		ds_{Dp}^2 &= \left(\frac{\rho_p}{u} \right)^{\frac{7-p}{2}} 
		\left[ -\frac{2 \tilde \Delta^2}{u^2} dt^2 - 2 dt d\xi + \sum_{i=1}^{p-1} dx_i^2 \right] 
		+ \left(\frac{\rho_p}{u} \right)^{\frac{p+1}{2}} \left[du^2 + u^2 d \Omega_{8-p}^2 \right] \;, \\
		e^{\Phi} &= \left(\frac{\rho_p}{u} \right)^{\frac{(p-3)(7-p)}{4}} \;, \qquad 
		B = \frac{\sqrt{2} \tilde \Delta \rho_p^2}{u^2} (-d\xi + A) \wedge dt  \;.
	\end{align}
where $\tilde \Delta$ can be eliminated by redefinition of $t$ and $\xi$. 
Note that $p=d+1$ due to the fact that one of the spatial coordinates combined into $\xi$ coordinate. 
We would like to compactify this theory on $S^{8-p}$ and show that it leads to hyperscaling violation. 

Dimensional reduction and going to Einstein frame give 
	\begin{align} 
		ds_{Dp}^2 &= r^{-\frac{2(9-p)}{p(5-p)}} 
		\left[ -\beta r^{-\frac{4}{5-p}} dt^2 - 2 dt d\xi + \sum_{i=1}^{p-1} dx_i^2 + dr^2 \right]  \;,
	\end{align}
where we used $ u = r^{\frac{2}{5-p}}$ and $\beta \propto \tilde \Delta^2$. 
Thus the non-relativistic D$p$ brane solutions (\ref{nonrelBrane}) 
gives the Schr\"odinger type theories with hyperscaling violation with the identifications  
	\begin{align}
		\frac{\theta}{D} = 1 - \frac{9-p}{p(5-p)} \;, \qquad z = \frac{7-p}{5-p} \;,
	\end{align}
where $p=d+1$ in our notation. Thus this general solution is defined for $p \leq 7$, and null energy condition 
(\ref{nullECondition}) is satisfied $p \leq 8$ covering the range. 
The hyperscaling violation exponent $\theta$ is zero for $p = 3$, negative for $p < 5$ and positive for $p > 5$. 
For $p > 5$, the dynamical exponent $z$ is negative, and thus they might not be physically interesting. 
Thus the dimensional reduction of D$p$ brane solutions gives only the restricted classes of metrics, 
$ \theta<0$ and $z>0 $ or $ \theta>0$ and $z<0 $, which seems to be interesting.  

In this case, the scalar equation of motion (\ref{scalarFieldEqMom}) can be written 
	\begin{align}     \label{scalarFieldEqMomDp}
		\left(\partial_r^2 - \frac{9-p}{(5-p)~ r}\partial_r - \vec k^2+ 2 M \omega - \beta \frac{M^2}{ r^{4/(5-p)}} 
		-\frac{m^2}{r^{2(9-p)/p(5-p)}}\right)\phi=0 \;,
	\end{align}    
For $p <5$, one of the last two terms in (\ref{scalarFieldEqMomDp}) is dominant at the boundary except 
$p=3$, and it is not clear how to evaluate the full correlation functions. 
The case $p=3$ ($d=2$) corresponds to the well known conformal Schr\"odinger case. 
It will be interesting to find ways to construct correlation functions from this low energy point of view. 

Entanglement entropy can be evaluated using (\ref{entanglementEntropyHyper}),  
and is given by 
	\begin{align}      \label{entanglementEntropyHyperDp}
		\mathcal  S_{Dp} 
		&= \frac{(R M_{Pl})^{p} }{4 (\alpha-1)} 
		\left(  \left(\frac{\epsilon}{R_\theta}\right)^{-\frac{(3-p)^2}{5-p}} \frac{L^{p-2} L_\xi }
		{\epsilon^{-\frac{p^2-6p+7}{5-p}}} 
		-  c_{\theta} ~\left(\frac{l}{R_\theta}\right)^{-\frac{(3-p)^2}{5-p}} \frac{L^{p-2} L_\xi }
		{l^{-\frac{p^2-6p+7}{5-p}}}  \right) \;,
	\end{align}   
where $\alpha, c_\theta$ are given in (\ref{entanglementEntropyHyper}) with $d=p-1$.
For $p=3 (d=2)$, using (\ref{entanglementEntropySchr}), we get 
	\begin{align}      
		\mathcal  S_{D3} 
		&= \frac{(R M_{Pl})^{3} }{4~ M_\xi} ~
		\left(  \frac{L  }{\epsilon} -  c_{2} \frac{L }{l}  \right) \;, \qquad 
		c_2 = \left( \frac{\sqrt{\pi} \Gamma\left(3/4 \right)}{\Gamma\left(1/4 \right)} \right)^2
	\end{align}   
which reveals area law of the entanglement entropy. 
	
From these discussions, it is clear that dimensional reduction of the non-relativistic D$p$ brane 
solutions \cite{topdownSchrMetric} do not provide Schr\"odinger type theories with the logarithmic 
violation of the entanglement entropy, signalling a Fermi surface of the dual field theory. 
There are other types of non-relativistic brane solutions listed in \cite{topdownSchrMetric},  
which seem to be interesting for further investigations.%
\footnote{Note that there are similar solutions, generated via 
Melvin twist in different directions \cite{PFT}, which are expected to 
give Lifshitz type theories with hyperscaling violation upon dimensional reduction. It will be interesting 
to check whether these solutions have more interesting properties such as Fermi surfaces.   
See earlier entanglement entropy calculations in similar backgrounds in \cite{Barbon:2008ut}. 
}

\section{Comments on $\beta=0$ case}        \label{sec:AdSLC}

Until now we concentrated on $\beta \neq 0$ case, which was considered in \cite{son}. 
There exist another viable candidate for the geometric realization of the Schr\"odinger holography, 
called ``AdS in light-cone frame'' (ALCF) \cite{goldberger} for the case $\beta=0$.%
\footnote{
See also \cite{Maldacena:2008wh}\cite{Kim:2010tf} for its finite temperature generalizations 
for $z=2$ case, which has extra parameter $b$ (mass dimension -1) 
to ensure correct dimensions for the $t$ and $\xi$ coordinates. 
It turns out that this theory has very interesting phenomenological magneto-transport properties, 
which show same universal features of very low temperature normal state of high $T_c$ cuprates 
superconductors \cite{KKP}. 
}
In this section, we would like to survey the similarities and differences of the AdS in light-cone 
compared to the Schr\"odinger background studied until now. 

The metric is given by 
	\begin{align} \label{zeroTAdSinLCMetric}
		ds^2 = r^{-2 + 2  \theta/D} \left(- 2 dt d\xi + \sum_{i=1}^{d} d x_i^2 + dr^2  \right) \;. 
	\end{align}
This metric with $\theta=0$ is simple and its finite temperature case is much more tractable than the case with $\beta \neq 0$. 
It turns out that holographic renormalization of this metric with $\theta =0$ is no more difficult than 
the corresponding AdS metric, and thermodynamic and transport properties are analyzed rigorously in 
\cite{Kim:2010tf}, even $r_c \rightarrow 0$ limit.

Before starting detailed analysis, we would like to comment crucial differences of the metric 
(\ref{zeroTAdSinLCMetric}) compared to (\ref{zeroTMetric}) and Lifshitz metric. 
First of all, even though we consider general dynamical exponent $z$, the ALCF metric (\ref{zeroTAdSinLCMetric}) 
does not show explicit dependance on $z$. 
Motivated by the discussions in \S \ref{sec:EntanglementEntropy}, we rewrite the metric 
(\ref{zeroTAdSinLCMetric}) as  
	\begin{align} \label{zeroTAdSinLCMetric2}
		ds^2 = r^{-2 + 2  \theta/D} \left(- 2 ( r^{1-z} dt) ( r^{z-1} d\xi ) + \sum_{i=1}^{d} d x_i^2 + dr^2  \right) \;. 
	\end{align}
Even though this seems to be an arbitrary splitting of $r$ dependence between the coordinate $t$ and $\xi$, 
this is a unique splitting to make the combination $r^{1-z} dt$ and $ r^{z-1} d\xi$ to have the same dimension 
as $x_i$ and $r$. 
It is well motivated for us to investigate various physical details of ALCF compared to those of the Schr\"odinger 
background in ADM form (\ref{ADMzeroTMetric}). Now explicit $z$ dependence can be visualized.  

For $\theta=0$, the metric is invariant under the translations, rotations and Galilean boost. 
Furthermore, it is also invariant under the special conformal transformation (\ref{SCT}) 
as well as the scaling transformation (\ref{ST}). Thus it has conformal Schr\"odinger symmetry 
for general $z$. For $\theta\neq 0$, the metric transforms covariantly 
for as $ ds \rightarrow \lambda^{\theta/D} ds$ and  $ds \rightarrow \left( \frac{r}{1+ct}\right)^{\theta/D} ds$ 
under the scaling (\ref{ST}) and conformal transformation (\ref{SCT}).

\subsection{Basic properties} 

We would like to do parallel analysis done in \S \ref{sec:basics}. 

\bigskip \bigskip
\noindent {\it Null energy condition} 

The Ricci tensor and scalar curvature for the metric (\ref{zeroTAdSinLCMetric}) are given by
	\begin{align} 
		R_{ij}&= - R_{t\xi} = - \frac{(D-\theta) (D(d+2) - (d+1)\theta)}{D^2 r^2} \;,  \nonumber\\
		R_{rr}&=\frac{(d+2)( \theta-D) }{D r^2} \;, \\ 
			\mathcal R ~~&= r^{-2\theta/D} \frac{(d+2)(\theta-D) (D(d+3) - (d+1)\theta)}{D^2} \;.
	\end{align}
The scalar curvature is then $\mathcal R \propto r^{-2 \theta /D}$, which becomes constant for $\theta=0$ as expected 
from the observation that the metric (\ref{zeroTAdSinLCMetric}) is conformally equivalent to ALCF metric. 
Energy momentum tensor can be computed as $T_{\mu\nu} = R_{\mu\nu} - 1/2 g_{\mu\nu} \mathcal R$.  

To consider various physically sensible dual field theories, 
we would like to constrain the parameters using null energy condition 
	\begin{align}   
	T_{\mu\nu} N^\mu N^\nu \geq 0 \;,
	\end{align}   
where the null vectors satisfy $N^\mu N_\mu = 0$. The two independent null vectors are 
	\begin{align}   
	N^t= \frac{1}{\sqrt{2}~ r^{\theta /D - z }} \;, \quad
	N^\xi = \frac{1}{\sqrt{2} ~r^{\theta /D + z -2}} \;, \quad
	N^r= \frac{\cos (\phi )}{r^{-1 +\theta /D}} \;, 
	\quad N^i= \frac{ \sin (\phi) }{r^{-1 +\theta /D}}  \;, 
	\end{align}   
where $\phi=0$ or $\pi/2$. 
Seemingly there are two independent conditions, but one is trivial. Thus we have  
	\begin{align}   \label{nullEConditionAdSLC}
		& \theta (\theta - D) \geq 0  \qquad \rightarrow \qquad  
		\theta \leq 0 \quad \text{or} \quad \theta \geq  D  \;.
	\end{align}
Note that it is independent of $z$.   
This is a similar condition obtained in \cite{Dong:2012se}, where Lorentz invariant case $z=1$ gives the 
same null energy condition. ALCF is directly derived from AdS metric and thus it is not surprising 
to have the same condition. 
Both ranges are realized in the string theory constructions for Lifshitz case \cite{Dong:2012se}.

\bigskip \bigskip 
\noindent {\it Holographic energy-momentum tensor}

Following the prescription given in (\ref{QLEPrescription}), we calculate the 
holographic one-point function of the energy-momentum tensor for the metric (\ref{zeroTAdSinLCMetric}). 
At the hypersurface $r=r_c$, the induced metric has the form 
	\begin{align}     
		\gamma_{ab} dx^a dx^b =r_c^{-2 + 2\theta/D} \left(- 2 dt d\xi 
		+ \sum_{i=1}^{d} dx_i^2 \right) \;.
	\end{align}     
The result is  
	\begin{align}     
		&\langle \hat \tau_{t\xi}\rangle =\langle \hat \tau_{\xi t}\rangle = 
			-\frac{(d+1)(1- \theta/D)}{r_c^{d+2-(d+1)\theta/D}} h_{t\xi} \;,  \\ 
		&\langle \hat \tau_{ij}\rangle = -\frac{(d+1)(1- \theta/D)}{r_c^{d+2-(d+1)\theta/D}}  h_{ij} \;.
	\end{align}   
From these expressions, we observe that the hyperscaling violation has the same effect of modifying 
the one-point function of the energy-momentum tensor compared to $\beta \neq 0$ case. 
Note that there is no $ \langle \hat \tau_{tt}\rangle$ component. 
The shift is proportional to $\theta$ and reduces to $\theta$ for $D=d+1$.

\subsection{Semiclassical propagators}       \label{semiclassicalPropagatorAdSLC} 

The action of the particle moving in the background (\ref{zeroTAdSinLCMetric}) is described by 
\begin{align}     \label{particleActionAdSLC}
S = - m \int d \lambda ~ r^{-1 + \theta/D} \sqrt{ - 2  \frac{dt}{d\lambda} \frac{d\xi}{d\lambda} 
+ \left(\frac{d r}{d\lambda}\right)^2  + \left(\frac{d x}{d\lambda}\right)^2} \;. 
\end{align}      
where $\lambda$ is the worldline coordinate. The notations are the same as in \S \ref{sec:basics}. 
The propagator between two points $x$ and $x'$ on a fixed radius $r = \epsilon$ is 
$G_\epsilon (x',x) \sim \exp (S(x',x))$. 

The action (\ref{particleActionAdSLC}) has differences in time direction compared to (\ref{particleAction}). 
Thus the semiclassical propagator of the static case is the same as that of $\beta \neq 0$, 
and we get the total geodesic distance given in (\ref{staticAction}).  
The propagator is again given by (\ref{staticPropagator}). 

\subsubsection{Timelike case}

The differences come from the timelike geodesics. For $\lambda =r$ and $\Delta x_i=0$, 
the action (\ref{particleActionAdSLC}) gives 
	\begin{align}    
		S = - m \int dr \; r^{-(D-\theta)/D}\sqrt{1 - 2 \dot t \dot \xi }\,.
	\end{align}     
	Due to the conservations along $t$ and $\xi$,  there are corresponding constants of motion  
	\begin{align}
		\Pi_\xi = r^{-(D-\theta)/D}  \dot t / \sqrt{1 - 2 \dot t \dot \xi } \;, \qquad \quad 
		\Pi_t = r^{-(D-\theta)/D}  \dot \xi / \sqrt{1 - 2 \dot t \dot \xi } \;,
	\end{align}	
from which we get $\dot \xi = \left(  \Pi_t / \Pi_\xi  \right) \dot t$. 
Solving these two equations, we get the expressions for $\dot t$ and $\dot \xi$. 
	\begin{align}      \label{timelikeDottEqAdSLC}
		&\dot t = \Pi_\xi / \sqrt{2 \Pi_t \Pi_\xi +  r^{-2\frac{D-\theta}{D}} } \;,   \qquad
		\dot \xi =  \Pi_t / \sqrt{2 \Pi_t \Pi_\xi +  r^{-2\frac{D-\theta}{D}} }  \;.
	\end{align}
Using the boundary condition for the $t$ coordinate, $dr / dt |_{r=r_t}=0$ at the turning point, 
we fix one of the two constants as
	\begin{align}
		2  \Pi_t \Pi_\xi  = - r_t^{-2(D-\theta)/D} \;. 
	\end{align}
Using this we can rewrite (\ref{timelikeDottEqAdSLC}) as 
	\begin{align}  
		&\frac{d t}{dr} = \Pi_\xi \int \frac{r^{(D-\theta)/D}}{\sqrt{1-  (r/r_t)^{2(D-\theta)/D} }}  \quad \rightarrow \quad
		\frac{|\Delta t|}{2} =  \Pi_\xi r_t^{2-\theta/D} \frac{\sqrt{\pi} \Gamma \left(\frac{2D-\theta}{2(D-\theta)} \right)}{\Gamma \left(\frac{D}{2(D-\theta)} \right)}\;.
	\end{align}
And the action has the following form 
	\begin{align}    \label{timelikeGActionAdSLC}
		S = -2 m \int_\epsilon^{r_t} dr ~ \frac{r^{-(D-\theta)/D}}{ \sqrt{1-  (r/r_t)^{2(D-\theta)/D}  }}
		 = m \frac{2 D}{\theta} \epsilon^{\theta/D} - m \frac{2 D}{\theta} c_{\xi}~ 
			|\Delta t|^{\theta/(2D-\theta)} \;.
	\end{align}   
This result is the same as the case evaluated in (\ref{timelikeActionResultPsmall}) and $c_\xi$ is given there. 
The corresponding propagator can be obtained by exponentiating this action. 
	\begin{align}    \label{timelikePropagatorSmallP2AdSLC}
		G(\Delta t) \sim  \exp \left[2m \frac{D}{\theta} ~ \epsilon^{\theta/D}\right]  
		 \exp \left[-2 m \frac{D}{\theta} c_{\xi} ~|\Delta t|^{\frac{\theta}{(2D -\theta)}}\right] \;.
	\end{align}

\subsubsection{General case} 

General case also can be evaluated by considering the action 
	\begin{align}    
		S = - m \int d r \, r^{-(D-\theta)/D}\sqrt{ -2 \dot t \dot \xi + 1 + \dot x_i^2} \;, 
	\end{align}    
with $\lambda =r$.  
There are three constants of motion $\Pi_i, \Pi_t$ and $\Pi_\xi$, 
	The integrated $x, t, \xi$ equations of motion define three conserved momenta:
	\begin{gather}    
	\Pi_i =\frac{r^{-(D-\theta)/D} \dot x}{\sqrt{ -2 \dot t \dot \xi + 1 + \dot x^2}} \;,  \quad
	\Pi_t =\frac{- r^{-(D-\theta)/D}  \dot \xi }{\sqrt{-2 \dot t \dot \xi + 1 + \dot x^2}} \;, \quad 
	\Pi_\xi =\frac{-r^{-(D-\theta)/D} \dot t}{\sqrt{ -2 \dot t \dot \xi + 1 + \dot x^2}} \;.
	\end{gather}    
which can be solved 
	\begin{gather}    
		\frac{dx}{dr}=  \frac{- \Pi_i}{\sqrt{ 2 \Pi_t   \Pi_\xi - \Pi_i^2   + r^{ -2(D-\theta)/D}}} \;, \qquad 
		\frac{dt}{dr}=   \frac{\Pi_\xi}{\sqrt{ 2 \Pi_t   \Pi_\xi - \Pi_i^2   + r^{ -2(D-\theta)/D}}} \;, \\
		\frac{d\xi}{dr}=  \frac{\Pi_t }{\sqrt{ 2 \Pi_t   \Pi_\xi - \Pi_i^2  + r^{ -2(D-\theta)/D}}} \;.    \nonumber
	\end{gather}    
Also, using the fact that at the turning point $dr/dx |_{r=r_t} =0 , dr/dt |_{r=r_t}=0$, we can derive a relationship 
between $r_t,\Pi_i,\Pi_t$ and $\Pi_\xi$,
	\begin{align}      \label{canstraintGeneralPAdSLC}
		 2 \Pi_t  \Pi_\xi - \Pi_i^2    + r_t^{ -2(D-\theta)/D} =0 \;.
	\end{align}    
Plugging this into the geodesic equations, we get  
	\begin{gather}     \label{generalDotEqAdSLC} 
		\frac{dx}{dr}=  \frac{- \Pi_i ~ r^{ (D-\theta)/D}}{\sqrt{1- (r/r_t)^{ 2(D-\theta)/D}}} \;,  \quad 
		\frac{dt}{dr}=   \frac{\Pi_\xi~ r^{ (D-\theta)/D}}{\sqrt{1- (r/r_t)^{ 2(D-\theta)/D}}} \;, \quad
		\frac{d\xi}{dr}=  \frac{\Pi_t ~ r^{ (D-\theta)/D}}{\sqrt{1- (r/r_t)^{ 2(D-\theta)/D}}}  \;.  \nonumber
	\end{gather}    
The first two equations can be integrated to give 
	\begin{align}  
		\frac{|\Delta x_i|}{2} =- \Pi_i c_\theta r_t^{2-\theta/D}  \;, 
		\qquad 
		\frac{|\Delta t|}{2} = \Pi_\xi c_\theta r_t^{2-\theta/D}  \;, 
		\qquad 
		c_\theta =\frac{\sqrt{\pi} 
		\Gamma \left(\frac{2D-\theta}{2(D-\theta)}\right)}{\Gamma \left(\frac{D}{2(D-\theta)}\right)}  \;.
	\end{align} 
The expression for the total geodesic distance  
	\begin{align}   
	S = - m \int d r ~ \frac{r^{-(D-\theta)/D}}{\sqrt{ 1- (r/r_t)^{ 2(D-\theta)/D}}} 
		= m \frac{2 D}{\theta} \epsilon^{\theta/D} - m \frac{2 D}{\theta} ~c_\theta~  r_t^{\theta/D} \;.  
	\end{align}    

To get the expression for $r_t$ in terms of $r_t (|\Delta x_i|, |\Delta t|)$ and $\Pi_t$, 
we analyze the constraint equation (\ref{canstraintGeneralPAdSLC})
	\begin{align}    \label{reducedConstraintAdSLC} 
		& \frac{|\Delta t|}{c_{\theta}} \Pi_t ~ r_t^{\theta/D} - \frac{|\Delta x_i|^2}{4 c_\theta^2}
		 r_t^{-2+2 \theta/D}  + r_t^{ 2 \theta/D} = 0 \;, 
	\end{align}  
where we leave $\Pi_t$ undetermined because it is related to the motion along $\xi$. 

Instead of solving the equation (\ref{reducedConstraintAdSLC}) generally, 
we can consider three different cases for the (\ref{reducedConstraintAdSLC}). 
First, $\Delta t = 0$, $r_t = |\Delta x_i|/ (2 c_\theta)$, then we get 
	\begin{align}        \label{spacelikePropagatorGeneralAdSLC}
		G(\Delta x_i) \sim  \exp\left[m \frac{2 D}{\theta}~ \epsilon^{\theta/D}\right] 
		\exp\left[- m \frac{2D}{\theta} c_\theta  
		\left( \frac{|\Delta x_i|}{2 c_\theta }\right)^{\theta/D}   \right] \;.
	\end{align}     
The result is very similar to the static case we considered in (\ref{staticPropagator}).

Second, $\Delta x_i = 0$, $r_t^{\theta/D} = - \Pi_t |\Delta t|/ c_\theta$, then we get
	\begin{align}        \label{timelikePropagatorGeneralAdSLC}
		G(\Delta t) \sim  \exp\left[m \frac{2 D}{\theta}~ \epsilon^{\theta/D}\right] 
		\exp\left[- m \frac{2D}{\theta} (-\Pi_t) |\Delta t| \right] \;.
	\end{align}     
To make sense of this for $	|\Delta t| \gg 1$, we require $-\Pi_t > 0$. 
	
Third, for $|\Delta x_i|, |\Delta t| \gg 1$, the last term in (\ref{reducedConstraintAdSLC}) 
is negligible. $r_t$ is given by 
	\begin{align}   
	& r_t \approx  \left( \frac{1}{2 c_\theta \Pi_t} \frac{|\Delta x_i|^2}{2 |\Delta t|}\right)^{D/(2D-\theta)} \;.
	\end{align}  
Thus, the propagator for this particular timelike path is given by 
	\begin{align}        \label{specialPropagatorGeneralAdSLC}
	G(\Delta t, \Delta x_i) \sim  \exp\left[m \frac{2 D}{\theta}~ \epsilon^{\theta/D}\right] 
	\exp\left[- m \frac{2D}{\theta} c_\theta  
	\left( \frac{1}{4 c_\theta \Pi_t} \frac{|\Delta x_i|^2}{2  |\Delta t|}\right)^{\theta/(2D-\theta)}   \right] \;.
	\end{align}     
This propagator is similar to the non-relativistic Schr\"odinger propagator modified by the $\theta$ dependent 
power. Thus the propagator behaves as $G \sim \exp\left[- m \left(\frac{|\Delta x_i|^2}{2  |\Delta t|}\right)^{\theta/(2D-\theta)} \right]$. It is interesting to have this form in the semiclassical 
regime. This is similar to (\ref{timelikePropagatorSmallP}).

\subsection{Correlation functions}    \label{sec:correlationFunctionAdSLC}

In this section, we consider the correlation functions of the probe scalar similar to \S \ref{sec:correlationFunction}. 
The action of the scalar field is the same as (\ref{ScalarAction}), which is coupled to the metric 
(\ref{zeroTAdSinLCMetric}). 
The equation of motion for a scalar field with mass $m$ is given in the momentum space as 
	\begin{align}     \label{scalarFieldEqMomAdSLC}
		\left(\partial_r^2-\frac{(d+1)(D-\theta)}{D~ r}\partial_r - \vec k^2+ 2 M \omega 
		-\frac{m^2}{r^{2(D-\theta)/D}}\right)\phi=0 \;,
	\end{align}    
where $\vec k, \omega $ are Fourier transform of $\vec x, t$, respectively. We treat the $\xi$ direction 
special and replace $\partial_\xi = i M$ for the scalar field \cite{son}. Compared to (\ref{scalarFieldEqMom}), 
this equation is simpler. Note that this equation does not depend on the dynamical exponent $z$, 
and thus correlation functions do not depend on it explicitly. For the detailed analysis, 
only the last term in the equation needs care.  
For zero temperature background, we are concerned for both the boundary behavior at $r\to 0$ 
and the behavior deep in the bulk at $r\to \infty$ of the bulk field. 

Let us consider the $\phi$ at the boundary, $r \rightarrow 0$. 
For $\theta>0$ at small $r$, all the terms except the first two are subdominant. 
Thus we can solve the equation of motion at leading order in $r$ as $\phi\sim r^\nu$,
$\nu=0$ or $\nu = d+2-(d+1)\theta/D$. Following the prescription given around (\ref{CorrelationExpansion})
in \S \ref{sec:correlationFunction}, we can read off the momentum space correlation function.

\subsubsection{Schr\"odinger type for general $z$ with $\theta=0$}    \label{sec:corrAdSLC}

In this section we consider correlation functions of the Schr\"odinger-type theories {\it without} hyperscaling 
violation. The equation of motion reduces to 
	\begin{align}     \label{scalarFieldEqMomZerothetaAdSLC}
		\left(\partial_r^2-\frac{d+1}{ r}\partial_r - k^2  -\frac{m^2}{r^{2}}\right)\phi=0 \;,
	\end{align}    
Where $k^2 = \vec k^2 - 2 M \omega$. 
Note that, in this case, the term proportional to $m^2$ contribute as the same order as the derivative 
parts, and thus to the scaling dimension of the dual scalar operator. The large $m^2$ limit is corresponding 
to the large scaling dimension limit. 

There are several cases of interest, especially the cases $z=d+1$ and $z=d+2$ for the application of the 
Fermi surface and novel phases as we see in \S \ref{sec:EntanglementEntropyAdSLC}. 
For the analysis of the correlation functions for ALCF, these cases are no different from other values of $z$.  

The conformal dimension of the scalar operator is  
$\Delta = \frac{d+2}{2} \pm \nu=\frac{d+2}{2}\pm\sqrt{\left(\frac{d+2}{2}\right)^2+ m^2}$.
The solution that satisfies the proper boundary condition at $r=\infty$ is
	\begin{align}     \label{z=2SolAdSLC}
		\phi= (kr)^{1+d/2} K_{\nu}(kr) \;. 
	\end{align}     
Note that we have normalized the solution at the boundary according to \eqref{CorrelationExpansion}, and thus 
we find the momentum space correlation function as $G(k)\sim k^{2\nu}$, 
by expanding the modified Bessel function. 
Fourier transforming back to position space, we find the two-point function to be
	\begin{align}       \label{CFz=2AdSLC}
		\langle\mathcal O(x')\mathcal O(x)\rangle \sim \frac{\theta(\Delta t)}{|\Delta t|^\Delta} 
		e^{i M \frac{|\Delta \vec x|^2}{2 |\Delta t|}} \;,
	\end{align}      
where $\mathcal O$ is an operator dual to the massive $\phi$ in the bulk. 
This result is valid for all $z$.

\subsubsection{With hyperscaling violation, $\theta \neq 0$}   \label{sec:corrHSVAdSLC}

In this section we would like to consider the case with hyperscaling violation and to evaluate 
correlation functions satisfying the differential equation (\ref{scalarFieldEqMomAdSLC}).

\bigskip \bigskip 
\noindent {\it For $m^2 = 0$ case}

The equation of motion in momentum space is 
	\begin{align}     
		\left(\partial_r^2-\frac{(d+1)(D-\theta)}{D~ r}\partial_r - \vec k^2+ 2 M \omega \right)\phi=0\;.
	\end{align}    
This is the case we can see the effect of the hyperscaling violation clearly.  
The scaling dimension $\Delta_{\theta=0}= d+2$ of the scalar operator is shifted to  
$ \Delta =\Delta_{\theta=0} -\frac{(d+1)\theta}{D}$ due to the hyperscaling violation exponent $\theta$. 
The solution is given by Bessel function with explicit dependence of $\theta$. 
Following the prescription described above, we get an exact result 
	\begin{align}
		G(k)&= c_m  k^{\left(2+d-\frac{(1+d) \theta }{D}\right)} \;, \quad 
		c_m = 2^{-2-d+\frac{(1+d) \theta }{D}}  \frac{ \Gamma \left(- \frac{2+d}{2}+\frac{(1+d) \theta }
		{2D}  \right)}{\Gamma \left(\frac{2+d}{2}-\frac{(1+d) \theta }{2 D} \right)} \;,
	\end{align}
upto some numerical factor independent of momentum. 
This function can be exactly evaluated to give the position space correlation function as 
	\begin{align}     \label{corrZeroMassZequal1AdSLC}
		\langle\mathcal O(x')\mathcal O(x)\rangle 
		\sim \frac{\theta (\Delta t)}{|\Delta t|^{d+2-\frac{(d+1)\theta}{2D}}} 
		e^{i M \frac{|\Delta \vec x |^2}{2 |\Delta t|} }\;.
	\end{align}   
Thus we observe the effect of the hyperscaling violation exponent $\theta$.

\bigskip \bigskip 
\noindent {\it For $\theta = D$ case}

This case is particularly simple. The equation of motion has the form 
	\begin{align}   
		\left(\partial_r^2 - k^2 - m^2  \right)\phi=0\,.
	\end{align} 
Where $k^2 = \vec k^2 - 2 M \omega$. The solution has exponential form 
as $e^{\pm \sqrt{k^2 + m^2} r}$. Boundary condition picks up the negative sign and the solution, 
with a correct normalization, is 
	\begin{align}
		\phi = e^{- \sqrt{k^2   + m^2}~ r} \;.
	\end{align}
The combination $d+2-(d+1)\theta/D = 1$ for $\theta=D$. 
Thus two point function in momentum space is 
	\begin{align}
		G(k) = \sqrt{k^2 + m^2} \;,
	\end{align} 
which can be Fourier transform back to position space as 
	\begin{align}
		\langle\mathcal O(x')\mathcal O(x)\rangle 
		= \frac{1}{|\Delta t|^{\frac{d+3}{2}}} e^{i M \frac{|\Delta \vec x|^2}{2 |\Delta t|} 
		- i \frac{ m^2}{2M} |\Delta t|}\;.
	\end{align}     
Let us consider some special cases.  
At short distance, the two point function is dominated by the large $k$ behavior as 
$G(k)\sim k = \sqrt{\vec k^2 - 2M\omega}$, whose inverse Fourier transform gives 
	\begin{align}
		\langle\mathcal O(x')\mathcal O(x)\rangle 
		\sim \frac{1}{|\Delta t|^{\frac{d+3}{2}}} e^{i M \frac{|\Delta \vec x |^2}{2 |\Delta t|}}\;.
	\end{align}     
For $M\omega, M^2 \ll m^2, \vec k^2$, we can use the saddle point approximation to get 
	\begin{align}
		\langle\mathcal O(\vec x')\mathcal O(\vec x)\rangle 
		\sim e^{-m |\Delta \vec x|}\;.
	\end{align}     
This case reduces to (\ref{spacelikePropagatorGeneralAdSLC}) for $D=\theta$.

\bigskip \bigskip 
\noindent {\it For $\theta = \frac{d+1-z}{d+1}D$ case}

The case $\theta = \frac{d+1-z}{d+1}D$ is an interesting case from the point of view of entanglement entropy, 
where the logarithmic violation of the area law is observed below. Seemingly,  
the equation of motion explicitly depend on $z$ through the condition. 
\begin{align}     \label{scalarFieldEqMomLogViolationAdSLC}
\left(\partial_r^2-\frac{z}{r}\partial_r - k^2 
-\frac{m^2}{r^{2z/(d+1)}}\right)\phi=0 \;.
\end{align}  
But this is not the case. 
We can re-express the same condition as $z = (d+1)(D-\theta)/D$.   
Then the equation of motion goes back to the general one given in (\ref{scalarFieldEqMomAdSLC}).  
Thus for ALCF with $\beta = 0$, the distinction we see from the entanglement entropy analysis 
does not make differences. To progress further, we need to specify the parameters.

\bigskip  \bigskip 
\noindent {\it For $\theta = \frac{d+2-z}{d+1}D$ case}
 
There exists another interesting value of $\theta$, $\theta = \frac{d+2-z}{d+1}D$, where the entanglement entropy is 
proportional to the volume and the area law is extensively violated, which is observe below.  
In this case the equation of motion has the form 
	\begin{align}     \label{scalarFieldEqMomVolumeAdSLC}
		\left(\partial_r^2-\frac{z-1}{r}\partial_r - k^2 -\frac{m^2}{r^{2(z-1)/(d+1)}}\right)\phi=0 \;.
	\end{align}  
The explicit dependence of $z$ here is misleading. 
Using $z =1+ (d+1)(D-\theta)/D$, we can check that the equation reduces to (\ref{scalarFieldEqMomAdSLC}).  
The reason behind is that the equation (\ref{scalarFieldEqMomAdSLC}) we start with does not depend on 
$z$ and thus the condition $\theta = \frac{d+2-z}{d+1}D$ does not put any constraint on it.

\subsubsection{Scaling argument and summary} 

Let us consider the differential equation (\ref{scalarFieldEqMomAdSLC}) 
	\begin{align}     
		\left(\partial_r^2-\frac{(d+1)(D-\theta)}{D~ r}\partial_r - k^2 
		-\frac{m^2}{r^{2(D-\theta)/D}}\right)\phi=0  \;.
	\end{align}     
from different angle and with null energy condition, $\theta \leq 0$ or $\theta \geq D$.	
There are two special cases, $\theta=0$ and $\theta=D$, which are analyzed analytically in  
\S \ref{sec:corrAdSLC} and \S \ref{sec:corrHSVAdSLC}, respectively. 
The cases with $\theta < 0$ would be interesting to further investigate. 

For $\theta >D$, the mass term is not important at the boundary $r \rightarrow 0$ and thus we can use 
the prescription (\ref{CorrelationExpansion}). Then the naive scaling dimension is shifted to 
$\Delta= d+2-(d+1)\theta/D$. 
The equation (\ref{scalarFieldEqMomAdSLC}) is invariant under the Galilean boost, and the unique combination 
for energy and momenta $k^2 = \vec k^2-2 M \omega$ should be maintained. Moreover, 
there is a scaling symmetry in the equation
	\begin{align}     
	r\to\lambda r \;,\qquad
	k\to k/\lambda \;,\qquad
	m\to m/\lambda^{\theta/D} \;,
	\end{align}     
under which the coefficient function $G(k)$ should transform as 
$G(k ; m) = \lambda^{\Delta}~ G(k/\lambda ;  m/\lambda^{\theta/D} )$.
Then, the momentum space correlation function has the general form  
	\begin{align}
	G(k)\sim k^{\Delta} \cdot F (m/k^{\theta/D}) \;.
	\end{align}
When $m/k^{\theta/D} \ll 1$, $F$ is independent of $k$, and thus we can Fourier transform back to 
the position space to find 
	\begin{align}       
	\langle\mathcal O(x')\mathcal O(x)\rangle \sim 
	 \frac{\theta(\Delta t)}{|\Delta t|^{\Delta + (d+1) \theta / (2D)}} e^{i M \frac{|\Delta \vec x|^2}{2 |\Delta t|}} \;,
	\end{align}   
which is the general result with the effect of $\theta$.

\subsection{Entanglement Entropy}     \label{sec:EntanglementEntropyAdSLC}

From the discussion of \S \ref{sec:EntanglementEntropy}, it becomes clear that there is unique 
prescription for the minimal surface, and static and stationary cases end up to have the same result. 
Once we have this picture in mind, it is not difficult to guess that the result would be the same 
for ALCF. The form of the metric (\ref{zeroTAdSinLCMetric2}) is uniquely fixed if we use the power of 
$r$ to make up the dimension of $\xi$ to physical length. 

Using the prescription developed in \S \ref{sec:EntanglementEntropy}, we would like to 
compute the entanglement entropy for a strip with (non-)compact $\xi$ direction 
\begin{align}
0 \le \xi \le L_\xi\;,\quad -l \le x_1 \le l\;,\quad 0 \le x_i \le L\;,\;\quad i = 2,\cdots \;, d
\end{align}   
in the limit $l \ll L, L_\xi$. 
The strip is located at $r=\epsilon$, and the profile of the surface in the bulk is given by $r=r(x_1)$. 
Thus the area is given by 
\begin{align} 
\mathcal A = L^{d-1} L_\xi \int_0^{r_t} dr e^{(d+1) A(r) -B(r) }\sqrt{1+ \left(\frac{dx_1}{dr} \right)^2} \;,
\end{align}   
where we use $x_1=x_1(r)$ and $dr/dx_1|_{r_t}=0$. And 
\begin{align}
e^{A(r)} = r^{-1 + \theta/D} \;, \qquad  e^{B(r)} = r^{-z +1} \;. 
\end{align}
Thus there is no difference for the entanglement entropy of the ALCF compared to that of the Schr\"odinger case. 

\begin{figure}[!ht]
\begin{center}
	 \includegraphics[width=0.45\textwidth]{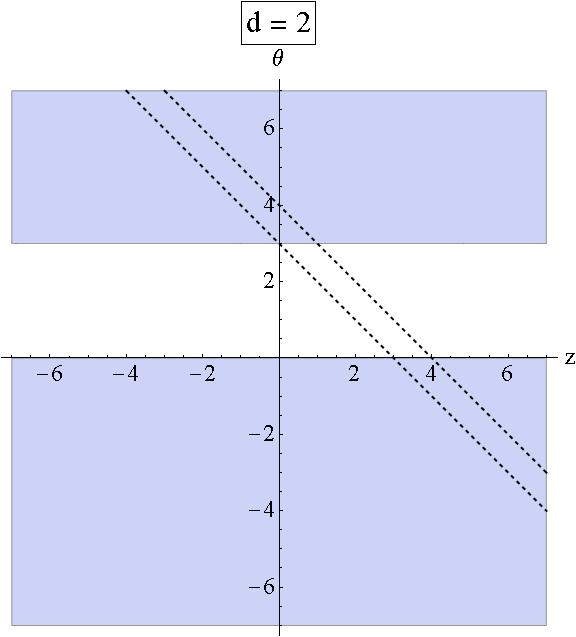} \quad 
	 \includegraphics[width=0.45\textwidth]{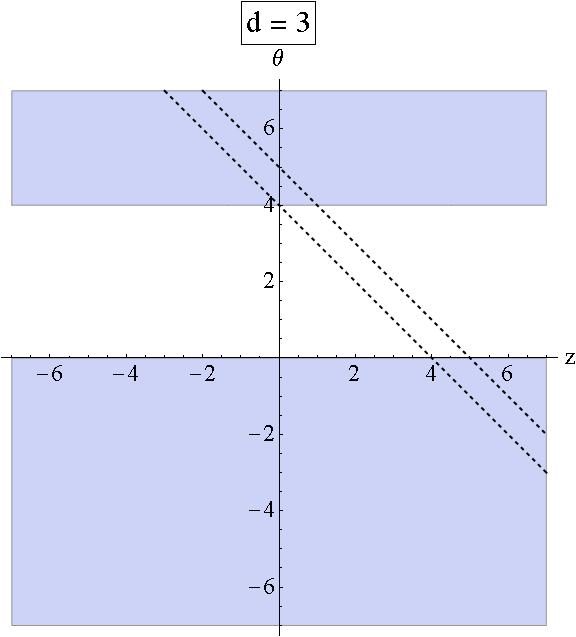}	 
	 \caption{The parameter ranges of $(z, \theta)$ for the novel phases in the case of ACLF with $\beta=0$ 
	 are plotted for $d=2$ and $d=3$. 
	 The plot assumes $D=d+1$. The novel phases lie in the region between the black dashed lines. 
	 The blue background is allowed regions from the null energy condition.  
	 }
	 \label{fig:NovelPhaseAllowedRegionsAdSLC}
\end{center}
\end{figure}

In summary, for $\theta=0$, the entanglement entropy is given by 
\begin{align}    \label{entanglementEntropyAdSLC}
\mathcal  S  
&= \frac{(R M_{Pl})^{(d+1)}}{4  (d-z+1)}  \left( \left(\frac{L}{\epsilon} \right)^{d-1}  
\left(\frac{L_\xi}{\epsilon^{2-z}} \right)
  -  c_z ~  \left(\frac{L}{l} \right)^{d-1}  \left(\frac{L_\xi}{l^{2-z}} \right)  \right) \;,
\end{align}   
where $c_z$ is given in (\ref{entanglementEntropySchr}). 
There exist also the logarithmic violation of the area law for $z=d+1$ as well as extensive violation for 
$z=d+2$. In between, there exist new novel phases, violating area law of entanglement entropy.  
For $\theta \neq 0$, we get the same entanglement entropy as (\ref{entanglementEntropyHyper}) 
\begin{align}      \label{entanglementEntropyHyperAdSLC}
\mathcal  S 
&= \frac{(R M_{Pl})^{(d+1)} }{4 (\alpha-1)} 
\left(  \left(\frac{\epsilon}{R_\theta}\right)^{(d+1)\theta/D} \frac{L^{d-1} L_\xi }{\epsilon^{d-z+1}} 
-  c_{\theta} ~\left(\frac{l}{R_\theta}\right)^{(d+1)\theta/D} \frac{L^{d-1} L_\xi }{l^{d-z+1}}  \right) \;,
\end{align}   
There exist also some parameter range, $\frac{d+1-z}{d+1} D < \theta < \frac{d+2-z}{d+1} D$, where 
the area law is violated. Thus it is expected to have some novel phases in the range, which is depicted in 
figure \ref{fig:NovelPhaseAllowedRegionsAdSLC}.

\section{Outlook}      \label{sec:outlook}

In this paper we considered various properties of Schr\"odinger-type holographic theories for general 
dynamical exponent $z$ with or without hyperscaling violation exponent $\theta$. The main results are 
summarized in \S \ref{sec:introduction}. We would like to conclude with some speculations for the future 
directions.   

First, it will be interesting to generalize our discussions to the finite temperature black hole solutions. 
For Lifshitz theories, the black hole solutions with hyperscaling violation is already proposed and 
analyzed in \cite{Dong:2012se}. Constructing black hole solutions with hyperscaling violation 
is not simple for the Schr\"odinger type theories, at least for $\beta \neq 0$.    
For $z=2, \theta =0$ and $\beta \neq 0$, the solution is constructed and analyzed in \cite{Herzog:2008wg}
\cite{Maldacena:2008wh}, using a very nice solution generating technique called null Melvin twist \cite{nullMelvinTwist}. 
The resulting black hole solutions are complicated and computations of physical properties are difficult.  

As we mentioned already, there is simpler and viable construction for the Sch\"odinger holography, 
AdS in light-cone (ALCF) with $\beta=0$. The corresponding black hole solution is also constructed and analyzed 
in \cite{Maldacena:2008wh}\cite{Kim:2010tf}. It turns out that, at least for $z=2, \theta=0$, 
the thermodynamic properties of these two black hole solutions are identical \cite{Maldacena:2008wh}\cite{Kim:2010tf} 
and some transport properties are also identical if comparison is reliable \cite{Kim:2010tf}. 
Thus we naively expect that the black hole constructions for $\beta \neq 0$ and $\beta =0$ with non-zero $\theta$ 
would give the same thermodynamic properties. 
It will be interesting to check this explicitly \cite{kim}. 

From the discussion in \S \ref{sec:stringConst}, it is clear that there are several interesting cases with  
$\theta < 0$. For those cases, we are not able to evaluate correlation functions due to technical difficulties. 
The situation is similar to Lifshitz type theories \cite{Dong:2012se}, there dimensional reduction of D2 brane solution 
\cite{Itzhaki:1998dd} provides an example with $\theta =-1/3$, which corresponds an important 
example of the holographic application to condensed matter. It will be interesting to see progress along the line. 

{\bf Note added:} After submitting the first version to arXiv, there appeared \cite{Narayan:2012hk}\cite{Singh:2012un}, 
which have some overlap with \S \ref{sec:stringConst}. 

\section*{Acknowledgments}

We are indebted to C. Hoyos, S. Hyun, E. Kiritsis, Y. Oz, M. Rangamani, S. Sachdev and J. Sonnenschein 
for various discussions and valuable comments. We are grateful to C. Hoyos and J. Sonnenschein 
for their comments and various suggestions on the draft. We also thanks to the organizers of 
``The Third Indian-Israeli International Meeting on String Theory: Holography and its Applications,"
Jerusalem, February 1-8, 2012, where some of the results were presented, and the author received valuable comments on 
minimal surface prescriptions. 
We are supported in part by the Centre of Excellence supported by the Israel Science Foundation (grant number 1468/06).

\appendix

\section{Useful Formula}    \label{sec:formula}

\subsection{Integration}

We put some general integral expressions extensively used in \S \ref{sec:WKBPropagator}.  
In the same section, we consider only the case where parameter $b$ is positive. 
	\begin{align}
		&\int_0^{r_t} dr ~\frac{(r/r_t)^a}{\sqrt{1-(r/r_t)^b}} =
		\frac{\sqrt{\pi } r_t \Gamma \left(\frac{1+a}{b}\right)}{b \Gamma \left(\frac{1+ a}{b} + \frac{1}{2}\right)} 
		\;, \qquad \text{for} \qquad r_t >0 \;,~~ b>0 \;, \\
		&\int_0^{r_t} dr ~ \frac{(r/r_t)^a}{\sqrt{1-(r/r_t)^b}} = -i \frac{\sqrt{\pi } r_t 
		\Gamma \left(-\frac{1+a}{b} + \frac{1}{2}\right)}{b \Gamma \left(-\frac{1+ a}{b} + 1\right)} \;, \qquad 
		\text{for} \qquad r_t >0 \;, ~~ b<0 \;. 
	\end{align}
	
\subsection{Fourier transform} 

In this appendix, we summarize some of the Fourier transform used in the calculations. 
First we consider the case for function only for $k^2 = \vec k^2 - 2 M \omega$, 
	\begin{align}
		&\frac{\theta (t)}{ |t|^{x} } \exp \left( i M \frac{\sum_{i=1}^d  x_i^2}{2 |t|} \right)  \quad \nonumber \\
		&\qquad \Longleftrightarrow \quad \pi  i^{x-1+(-1)^d} 2^{\frac{d+4}{2}-x}  
		\Gamma \left(\frac{d+2}{2} -x \right) ~M^{1-x}~ (\vec k^2 - 2M\omega )^{x-\frac{d+2}{2}}\;. 
	\end{align}
It turns out that we have more involved expression and it involves with some combination like 
$k^2 + \beta M^2 + m^2$. This case can be also exactly evaluated as 
	\begin{align}
		&\frac{\theta (t)}{ |t|^{x} } \exp \left( i M \frac{\sum_{i=1}^d  x_i^2}{2 |t|} - i \frac{P}{2} |t| \right)  \quad 
		\nonumber \\
		&\qquad \Longleftrightarrow \quad \pi  i^{x-1+(-1)^d} 2^{\frac{d+4}{2}-x} 
		\Gamma \left(\frac{d+2}{2} -x \right)~ M^{1-x} ~(\vec k^2 - 2M\omega + M P )^{x-\frac{d+2}{2}} \;. 
	\end{align}
Thus for the massive case, one can use $P = m^2/M$.

\section{Metric Properties}        \label{sec:EinsteinTensorApp}

In the appendix, we summarize some results of the metric properties for the metric (\ref{generalzeroTMetric}) 
	\begin{align} 
		ds^2 = e^{2A(r)} \left( - \beta e^{2B(r)}  dt^2 - 2 dt d\xi + \sum_{i=1}^{d} d x_i^2 + dr^2  \right) \;. 
	\end{align} 
The Ricci tensors and scalar curvature for this metric are given by
\begin{align} 
R_{tt}&= \beta e^{2 B(r)} \left( (d+1) A'(r) (A'(r) + B'(r)) + 2 B'(r)^2 + A''(r) + B''(r)\right) \;, \nonumber\\
R_{ii}&= - R_{t\xi} = - \left( (d+1) A'(r)^2 + A''(r) \right)  \;,  \qquad \quad 
R_{rr}= - (d+2) A''(r) \;, \nonumber \\ 
\mathcal R ~~&= - e^{-2 A(r)} (d+2) \left( (d+1) A'(r)^2 + 2 A''(r) \right) \;.
\end{align}
This result reduces to the (\ref{RicciTensor}) for $e^{2 A(r)} = r^{-2 + 2\theta/D}$ and $e^{2 B(r)} = r^{-2 z+ 2}$. 
And the corresponding Einstein tensors are 
\begin{align} 
G_{tt}&= - \beta e^{2 B(r)} \bigg( \frac{d (d+1)}{2} A'(r)^2 - (d+1) A'(r) B'(r) 
- 2 B'(r)^2 + (d+1) A''(r) - B''(r) \bigg) \;,    \nonumber\\
G_{ii}&= - R_{t\xi} =  \frac{d (d+1)}{2} A'(r)^2 + (d+1)  A''(r)   \;, \qquad \quad 
G_{rr}=  \frac{(d+1)(d+2)}{2} A''(r)  \;.
\end{align}
These results are also applied to (\ref{zeroTAdSinLCMetric}) by imposing $\beta =0$.

\end{document}